\documentclass[preprint,prd,nofootinbib,tightenlines,amsmath]{revtex4}
\usepackage{graphicx}
\usepackage{bm}
\usepackage{multirow}

\begin{document}
\baselineskip=15pt \parskip=4pt

\vspace*{3em}

\title{Constraints on a New Light Spin-One Particle \\ from Rare $\bm{b\to s}$ Transitions}

\author{Sechul Oh\vspace{-2ex}}
\affiliation{Institute of Physics, Academia Sinica, \\ Taipei 115, Taiwan}

\author{Jusak Tandean\vspace{-2ex}}
\affiliation{Center for Mathematics and Theoretical Physics and Department of Physics, \\
National Central University, Chungli 320, Taiwan \\
$\vphantom{\bigg|_{\bigg|}^|}$}


\begin{abstract}
The anomalously large like-sign dimuon charge asymmetry in semileptonic $b$-hadron
decays recently measured by the D0 Collaboration may be hinting at the presence of
$CP$-violating new physics in the mixing of $B_s$ mesons.
It has been suggested that the effect of a nonstandard spin-1 particle lighter than the $b$~quark
with flavor-changing couplings to $b$ and $s$ quarks can reproduce the D0 result
within its one-sigma range.
Here we explore the possibility that the new particle also couples to charged leptons
\,$\ell=e,\mu$\, and thus contributes to rare \,$b\to s$\, processes involving
the leptons.
We consider in particular constraints on its couplings from existing
experimental data on the inclusive \,$B\to X_s\ell^+\ell^-$\, and exclusive
\,$B\to K^{(*)}\ell^+\ell^-$\, decays,
as well as the anomalous magnetic moments of the leptons.
We find that there is parameter space of the particle that is allowed by the current data.
Future measurements of these $B$ transitions and rare decays of the $B_s$ meson, such as
\,$B_s\to(\phi,\eta,\eta')\ell^+\ell^-$\, and \,$B_s\to\ell^+\ell^-$,\,
at LHCb and next-generation $B$ factories can probe its presence or couplings more stringently.
\end{abstract}

\maketitle

\section{Introduction}

The D0 Collaboration has recently reported a new measurement of the like-sign
dimuon charge asymmetry in semileptonic $b$-hadron decays~\cite{Abazov:2010hv}
which disagrees with the standard model (SM) prediction by about three standard deviations.
Although this finding still needs to be confirmed by future experiments, it might be
a clue to the presence of unexpectedly substantial $CP$-violating new physics in
the mixing of $B_s$ mesons.
Subsequently, we have suggested as one of the possible scenarios for the new physics that
the D0 result could be attributed to a nonstandard spin-1 particle lighter than the $b$ quark,
with flavor-changing couplings to the $b$ and $s$ quarks~\cite{Oh:2010vc}.
Specifically, we showed that the effect of the new particle can lead to a prediction
which is consistent with the new data within its one-sigma range.

New-physics scenarios involving nonstandard spin-1 particles with masses of a few GeV or less
have been discussed to some extent in various other contexts in the literature.
Their existence is generally still compatible with currently available data and also desirable,
as they may offer possible explanations for some of the recent experimental anomalies and
unexpected observations.
For instance, a spin-1 boson having mass of a few GeV and couplings to both quarks and
leptons has been proposed to explain the measured value of the muon $g$$-$2 and the NuTeV anomaly
simultaneously~\cite{Gninenko:2001hx}.
As another example, ${\cal O}$(MeV) spin-1 bosons which can interact with dark matter as well
as leptons may be responsible for the observed 511-keV emission from the galactic
bulge~\cite{Hooper:2007jr}.
If its mass is at the GeV level, such a particle may be associated with the unexpected excess
of positrons recently observed in cosmic rays, possibly caused by dark-matter
annihilation~\cite{Foot:1994vd}.
In the context of hyperon decay, a~spin-1 boson with mass around 0.2\,GeV,
flavor-changing couplings to quarks, and a dominant decay mode into $\mu^+\mu^-$
can explain the three anomalous events of \,$\Sigma^+\to p\mu^+\mu^-$\,
reported by the HyperCP experiment several years ago~\cite{He:2005we,Chen:2007uv,Oh:2009fm}.
Although in these few examples the spin-1 particles tend to have suppressed couplings to SM
particles, it is possible to test their presence in future high-precision
experiments~\cite{Hooper:2007jr,Foot:1994vd,He:2005we,Chen:2007uv,Oh:2009fm,Pospelov:2008zw}.
It~is therefore also of interest to see if the light spin-1 boson that could be responsible for
the anomalous D0 result contributes to some other $b$-meson
processes, perhaps with detectable effects, which we will attempt to do in this paper.

Here we consider the possibility that the new spin-1 particle, which we shall
refer to as~$X$, has flavor-conserving couplings to the electron and muon,
besides its flavor-changing couplings to the $b$ and $s$ quarks.
Accordingly, it can contribute to a number of rare \,$b\to s$\, transitions involving
the charged leptons, via the quark-level process \,$b\to s X^*\to s\ell^+\ell^-$,\,
where \,$\ell=e,\mu$.\,
In particular, we will deal with the impact of $X$ on the inclusive $b$-meson decay
\,$\bar B\to X_s\ell^+\ell^-$\, and the exclusive ones
\,$\bar B\to\bar K^{(*)}\ell^+\ell^-$\, and \,$\bar B_s^{}\to\phi\ell^+\ell^-$,\,
all of which have been
observed~\cite{Aubert:2004it,Iwasaki:2005sy,Aubert:2008ju,Wei:2009zv,Aaltonen:2011cn},
as well as the leptonic decay \,$\bar B_s^{}\to\ell^+\ell^-$.\,
Using the existing experimental information on the $b$-meson decays as well as the recent
data from~D0 on $B_s$ mixing, we will explore constraints on the couplings of~$X$.
Since $X$ has flavor-conserving interactions with the leptons and hence contributes to their
anomalous magnetic moments, we will also take their measured values into account.

In the following section, we write down the relevant Lagrangians and derive the amplitudes
for the processes of interest.
In Sec.~\ref{numbers}, we present our numerical results for the constraints on the $X$ couplings
and provide some predictions which may be tested at LHCb or future $B$ factories.
We conclude in Sec.~\ref{concl} and collect some of the formulas in appendixes.

\section{Interactions and amplitudes\label{int}}

We adopt a model-independent approach, assuming in addition that $X$ carries no color or
electric charge and its couplings to the fermions have both vector and axial-vector parts.
The Lagrangian for its flavor-changing interactions with the $b$ and $s$~quarks
is then~\cite{Oh:2010vc}
\begin{eqnarray} \label{LbsX}
{\cal L}_{bsX}^{} \,\,=\,\,
-\bar s\gamma_\mu^{}\bigl(g_{Vs}^{}-g_{As}^{}\gamma_5^{}\bigr)b\,X^\mu \,\,+\,\, {\rm H.c.} ~,
\end{eqnarray}
where $g_{Vs}^{}$ and $g_{As}^{}$ parametrize the vector and axial-vector couplings,
respectively, and in general can be complex, which would be new sources of $CP$ violation.
Moreover, the flavor-conserving couplings of $X$ with a charged lepton $\ell$ is described by
\begin{eqnarray}   \label{LlX}
{\cal L}_{\ell X}^{}  \,\,=\,\,
-\bar\ell\gamma^\mu\bigl(g_{V\ell}^{}-g_{A\ell}^{}\,\gamma_5^{}\bigr)\ell\,X_\mu^{} ~,
\end{eqnarray}
where $g_{V\ell}^{}$ and $g_{A\ell}^{}$ are real parameters because of the hermiticity
of~${\cal L}_{\ell X}$.
We will study only the \,$\ell=e$ and $\mu$\, cases due to lack of
the relevant data for \,$\ell=\tau$\, at present.
In the absence of model specifics, $g_{Ve,Ae}^{}$ and $g_{V\mu,A\mu}^{}$ are not necessarily
related.
In principle, $X$~can have additional interactions, flavor-conserving and/or flavor-violating,
with other fermions which are parametrized by more coupling constants.
We assume that these additional parameters already satisfy other experimental constraints to
which they are subject, but which we do not cover in this paper.

Together ${\cal L}_{bsX}^{}$ and ${\cal L}_{\ell X}^{}$ generate the contributions of $X$ to
the above-mentioned \,$b\to s$\, transitions.
For the inclusive decay \,$\bar B\to X_s\ell^+\ell^-$,\, the resulting amplitude is
\begin{eqnarray} \label{Hb2sX2sll}
{\cal M}_{b\to s\bar\ell\ell}^X &=&
-\frac{\bar s\gamma^\mu\bigl(g_{Vs}^{}-g_{As}^{}\gamma_5^{}\bigr)b\,
\bar \ell\gamma_\mu^{}\bigl(g_{V\ell}^{}-g_{A\ell}^{}\gamma_5^{}\bigr)\ell}
{q^2-m_X^2+i\Gamma_X^{}m_X^{}}
\nonumber\\ && \!\! - ~
\frac{2g_{A\ell}^{}\,m_\ell^{}\,\bar s\bigl[\bigl(m_b^{}-m_s^{}\bigr)g_{Vs}^{}
+ \bigl(m_b^{}+m_s^{}\bigr)g_{As}^{}\gamma_5^{}\bigr]b\,\bar\ell\gamma_5^{}\ell}
{m_X^2\bigl(q^2-m_X^2+i\Gamma_X^{}m_X^{}\bigr)} ~,
\end{eqnarray}
where \,$q=p_{\ell^+}^{}+p_{\ell^-}^{}$\, is the combined momentum of the dilepton and
$\Gamma_X^{}$ the total width of~$X$.
We remark that the presence of the $q^2$ dependence in the denominators distinguishes this
new-physics scenario from others involving heavy particles, which would induce four-fermion
operators with coefficients independent of $q^2$, as have been studied in
the literature~\cite{Greub:1994pi,Altmannshofer:2008dz}.
The SM counterpart of ${\cal M}_{b\to s\bar\ell\ell}^X$ is well known and
given by~\cite{Deshpande:1988mg}
\begin{eqnarray} \label{smHb2sll}
{\cal M}_{b\to s\bar\ell\ell}^{\rm SM} &=&
\frac{-\alpha G_{\rm F}^{}V_{ts}^*V_{tb}^{}}{\sqrt2\,\pi} \Biggl[
C_9^{\rm eff}\,\bar s\gamma^\mu P_{\rm L}^{}b\,\bar\ell\gamma_\mu^{}\ell \,+\,
C_{10}^{\rm eff}\,\bar s\gamma^\mu P_{\rm L}^{}b\,\bar\ell\gamma_\mu^{}\gamma_5^{}\ell
\nonumber \\ && \hspace*{14ex} -\;
\frac{2i C_7^{\rm eff}}{q^2}\,q^\nu\,\bar s\sigma^{\mu\nu}
\bigl(m_b^{}\,P_{\rm R}^{}+m_s^{}P_{\rm L}^{}\bigr)b\,
\bar\ell\gamma_\mu^{}\ell \Biggr] ~,
\end{eqnarray}
where \,$\alpha=e^2/(4\pi)$\, and $G_{\rm F}^{}$ denote the usual fine-structure and Fermi constants,
respectively, $V_{kl}$ are Cabibbo-Kobayashi-Maskawa matrix elements, $C_{7,9,10}^{\rm eff}$ are
Wilson coefficients evaluated at a~scale \,$\mu\sim m_b^{}$,\, and
\,$P_{\rm L,R}^{}=\frac{1}{2}(1\mp\gamma_5^{})$.\,
From the sum of the SM and $X$-induced amplitudes follows the decay rate
\begin{eqnarray} \label{Gb2sll}
\Gamma\bigl(\bar B\to X_s\ell^+\ell^-\bigr)  \,\,=\,\,
\Gamma_{b\to s\bar\ell\ell}^{\rm SM} \,+\, \Gamma_{b\to s\bar\ell\ell}^X ~,
\end{eqnarray}
where
\begin{eqnarray} \label{smGb2sll}
\Gamma_{b\to s\bar\ell\ell}^{\rm SM} &=&
\frac{\alpha^2G_{\rm F}^2\,|\lambda_t^{}|^2}{768\,\pi^5\,m_b^3}\int dq^2 \Biggl[
\bigl(|C_9^{\rm eff}|^2+|C_{10}^{\rm eff}|^2\bigr)\bigl(m_b^2+2q^2\bigr) +
\frac{4\bigl|C_7^{\rm eff}\bigr|^2m_b^2}{q^2}\bigl(2m_b^2+q^2\bigr)
\nonumber \\ && \hspace*{18ex} +\;
12\,{\rm Re}\bigl(C_9^{\rm eff*}C_7^{\rm eff}\bigr)m_b^2 \Biggr] \bigl(m_b^2-q^2\bigr)^2 ~,
\end{eqnarray}
\begin{eqnarray} \label{Gb2sX}
\Gamma_{b\to s\bar\ell\ell}^X &=&
\frac{\alpha^2G_{\rm F}^2\,|\lambda_t^{}|^2}{768\,\pi^5\,m_b^3} \int dq^2\, \biggl\{
{\rm Re}\biggl[\kappa\, \bigl(C_9^{\rm eff*}g_{V\ell}^{}-C_{10}^{\rm eff*}g_{A\ell}^{}\bigr)
\frac{g_{Vs}^{}+g_{As}^{}}{\Delta_X} \biggr] \bigl(m_b^2+2q^2\bigr)
\nonumber \\ && \hspace*{19ex} +\;
6\,{\rm Re} \biggl( \kappa\, C_7^{\rm eff*}\, \frac{g_{Vs}^{}+g_{As}^{}}{\Delta_X} \biggr)
g_{V\ell}^{}\,m_b^2 \biggr\} \bigl(m_b^2-q^2\bigr)^2
\nonumber \\ && +\;
\frac{\alpha^2G_{\rm F}^2\,|\lambda_t^{}|^2}{768\,\pi^5\,m_b^3}\, \frac{|\kappa|^2}{2}
\bigl(|g_{Vs}^{}|^2+|g_{As}^{}|^2\bigr)\bigl(g_{V\ell}^2+g_{A\ell}^2\bigr)\int dq^2\;
\frac{\bigl(m_b^2+2q^2\bigr)\bigl(m_b^2-q^2\bigr)^2}{|\Delta_X|^2} ~,
\end{eqnarray}
with $m_s^{}$ and $m_\ell^{}$ having been set to zero and
\begin{eqnarray}
\lambda_t^{} \,\,=\,\, V_{ts}^*V_{tb}^{} ~, \hspace{5ex}
\kappa \,\,=\,\, \frac{2\sqrt2\,\pi}{\alpha\,G_{\rm F}^{}\,\lambda_t^{}} ~, \hspace{5ex}
\Delta_X^{} \,\,=\,\, q^2-m_X^2+i\Gamma_X^{}m_X^{} ~.
\end{eqnarray}
Evidently $\Gamma_{b\to s\bar\ell\ell}^X$ contains not only the $X$-induced amplitude alone, but also its
interference with the SM one.

For \,$\bar B\to\bar K\ell^+\ell^-$,\, the amplitude follows from the effective Hamiltonians
${\cal H}_{b\to s\bar\ell\ell}^{{\rm SM},X}$ which yield Eqs.~(\ref{Hb2sX2sll}) and~(\ref{smHb2sll}).
Adding the $X$ and SM contributions, we can write it as
\begin{eqnarray} \label{MB2Kll}
{\cal M}\bigl(\bar B\to\bar K\ell^+\ell^-\bigr) \,\,=\,\,
\frac{-\alpha G_{\rm F}^{}\,\lambda_t^{}}{2\sqrt2\,\pi} \Bigl\{
A\, \bigl(p_B^{}+p_K^{}\bigr){}^\mu\, \bar\ell\gamma_\mu^{}\ell \,+\,
\bigl[ C\, \bigl(p_B^{}+p_K^{}\bigr){}^\mu + D\,q^\mu \bigr]
\bar\ell\gamma_\mu^{}\gamma_5^{}\ell \Bigr\} ~,
\end{eqnarray}
where  \,$q=p_{\ell^+}^{}+p_{\ell^-}^{}=p_B^{}-p_K^{}$,\,
\begin{eqnarray}
A &=& \biggl( C_9^{\rm eff} +
\frac{\kappa\, g_{Vs}^{}\, g_{V\ell}^{}}{\Delta_X^{}} \biggr) F_1^{} \,+\,
\frac{2m_b^{}\, C_7^{\rm eff} F_T^{}}{m_B^{}+m_K^{}} ~, \hspace{5ex}
C \,\,=\, \biggl( C_{10}^{\rm eff} -
\frac{\kappa\, g_{Vs}^{}\, g_{A\ell}^{}}{\Delta_X^{}} \biggr) F_1^{} ~,
\nonumber \\
D &=& C_{10}^{\rm eff}\; \frac{m_B^2-m_K^2}{q^2} \bigl(F_0^{}-F_1^{}\bigr) \,+\,
\frac{m_B^2-m_K^2}{m_X^2\,q^2}\;
\frac{\kappa\,g_{Vs}^{}\,g_{A\ell}^{}\,\bigl[F_1^{}\,m_X^2+F_0^{}\bigl(q^2-m_X^2\bigr)\bigr]}
{\Delta_X^{}} ~,
\end{eqnarray}
with $F_{0,1,T}^{}$ denoting the form factors for the \,$\bar B\to\bar K$\, matrix elements of
the \,$b\to s$\, quark operators in ${\cal H}_{b\to s\bar\ell\ell}^{{\rm SM},X}$
and being defined in Appendix~\ref{ff}.
One can see that the \,$\bar B\to\bar K\ell^+\ell^-$\, amplitude is independent of~$g_{As}^{}$.
The corresponding decay rate is given in Appendix~\ref{rates}.

It is worth mentioning here that the $B_s^{}$-meson decay  \,$B_s^{}\to P\ell^+\ell^-$\, involving
a pseudoscalar meson $P$ containing an $s\bar s$ component in its quark content,
such as $\eta$ and $\eta'$, also has an amplitude independent of~$g_{As}^{}$.\,
Consequently, the observation of \,$B_s^{}\to P\ell^+\ell^-$\, will provide additional
information on the $X$ contributions involving the products \,$g_{Vs}^{}\, g_{V\ell}^{}$\, and
\,$g_{Vs}^{}\, g_{A\ell}^{}$.\,

For \,$\bar B\to\bar K^*\ell^+\ell^-$,\, the amplitude from the SM and $X$ contributions
can be expressed as
\begin{eqnarray} \label{MB2Ksll}
{\cal M}\bigl(\bar B\to\bar K^*\ell^+\ell^-\bigr) &=&
\frac{-\alpha G_{\rm F}^{}\lambda_t^{}}{2\sqrt2\,\pi} \Bigl\{ \Bigl[
{\cal A}\,\epsilon_{\mu\nu\sigma\tau}^{}\,\varepsilon^{*\nu}p_B^\sigma p_{K^*}^\tau
- i{\cal C}\,\varepsilon_\mu^* +
i{\cal D}\,\varepsilon^{*}\!\cdot\!q\, (p_B^{}+p_{K^*}^{})_\mu^{}
\Bigr] \bar\ell\gamma^\mu\ell
\nonumber \\ && \hspace*{-2ex} +\; \Bigl[
{\cal E}\,\epsilon_{\mu\nu\sigma\tau}^{}\,\varepsilon^{*\nu}p_B^\sigma p_{K^*}^\tau
- i{\cal F}\,\varepsilon_\mu^* +
i{\cal G}\,\varepsilon^{*}\!\cdot\!q\, (p_B^{}+p_{K^*}^{})_\mu^{}
+ i{\cal H}\,\varepsilon^{*}\!\cdot\!q\,q_\mu^{}
\Bigr] \bar\ell\gamma^\mu\gamma_5^{}\ell \Bigr\} ~, ~~~~~
\end{eqnarray}
where \,$q=p_{\ell^+}^{}+p_{\ell^-}^{}=p_B^{}-p_{K^*}^{}$,\,
\begin{eqnarray} && \hspace*{5ex}
{\cal A} \,\,=\, \biggl( C_9^{\rm eff}+\frac{\kappa\,g_{Vs}^{}\,g_{V\ell}^{}}{\Delta_X} \biggr)
\frac{2\,V^{}}{m_B^{}+m_{K^*}^{}}+\frac{4 m_b^{}\,C_7^{\rm eff}\,T_1^{}}{q^2} ~,
\nonumber \\ &&
{\cal C} \,\,=\, \biggl( C_9^{\rm eff} +
\frac{\kappa\, g_{As}^{}\, g_{V\ell}^{}}{\Delta_X} \biggr)
A_1^{}\bigl(m_B^{}+m_{K^*}^{}\bigr)
+ 2m_b^{}\,C_7^{\rm eff}\,T_2^{}\,\frac{m_B^2-m_{K^*}^2}{q^2} ~,
\nonumber \\ &&
{\cal D} \,\,=\, \biggl( C_9^{\rm eff} +
\frac{\kappa\,g_{As}^{}\,g_{V\ell}^{}}{\Delta_X} \biggr)
\frac{A_2^{}}{m_B^{}+m_{K^*}^{}} +
2m_b^{}\,C_7^{\rm eff}\,\Biggl(\frac{T_2^{}}{q^2}+\frac{T_3^{}}{m_B^2-m_{K^*}^2}\Biggr) ~,
\nonumber \\
{\cal E} &=& \biggl( C_{10}^{\rm eff} -
\frac{\kappa\, g_{Vs}^{}\, g_{A\ell}^{}}{\Delta_X} \biggr)
\frac{2\, V^{}}{m_B^{}+m_{K^*}^{}} ~, \hspace{5ex}
{\cal F} \,\,=\, \biggl( C_{10}^{\rm eff} -
\frac{\kappa\, g_{As}^{}\,g_{A\ell}^{}}{\Delta_X} \biggr)
A_1^{}\bigl(m_B^{}+m_{K^*}^{}\bigr) ~,
\nonumber \\ && \hspace*{15ex}
{\cal G} \,\,=\, \biggl( C_{10}^{\rm eff} -
\frac{\kappa\, g_{As}^{}\,g_{A\ell}^{}}{\Delta_X} \biggr)
\frac{A_2^{}}{m_B^{}+m_{K^*}^{}} ~,
\nonumber \\
{\cal H} &=& \biggl( C_{10}^{\rm eff} -
\frac{\kappa\, g_{As}^{}\, g_{A\ell}^{}}{\Delta_X} \biggr)
\frac{\bigl(A_1^{}-A_2^{}\bigr)m_B^{}+\bigl(A_1^{}-2A_0^{}+A_2^{}\bigr)m_{K^*}^{}}{q^2}
\,-\, \frac{2\kappa\,g_{As}^{}\,g_{A\ell}^{}\, A_0^{}m_{K^*}^{}}{\Delta_X^{}\,m_X^2} ~, ~~~~
\end{eqnarray}
the form factors $V^{}$, $A_{0,1,2}^{}$, and $T_{1,2,3}^{}$ for
the \,$\bar B\to\bar K^*$\, transition being defined in Appendix~\ref{ff}.
The corresponding squared amplitude and decay rate are given in Appendix~\ref{rates}.

From the squared amplitude for \,$\bar B\to\bar K^*\ell^+\ell^-$,\, one can arrive at two
additional observables which have been measured besides the branching ratio.
They are the $\bar K^*$ longitudinal polarization fraction $F_L^{}$ and
the lepton forward-backward asymmetry $A_{\rm FB}^{}$, which are defined
from~\cite{Burdman:1995ks,Kruger:2005ep}
\begin{eqnarray} \label{d2GB2Ksll}
\frac{1}{d\Gamma\bigl(\bar B\to\bar K^*\ell^+\ell^-\bigr)/dq^2}
\frac{d^2\Gamma\bigl(\bar B\to\bar K^*\ell^+\ell^-\bigr)}{dq^2\,d(\cos\theta)} &=&
\frac{3}{4}\bigl(1-\cos^2\theta\bigr)F_L^{}
+ \frac{3}{8}\bigl(1+\cos^2\theta\bigr)\bigl(1-F_L^{}\bigr)
\nonumber \\ && +\; A_{\rm FB}^{}\,\cos\theta ~,
\end{eqnarray}
where $\theta$ is the angle between the directions of
$\bar B$ and $\ell^-$ in the dilepton rest frame.
Since these observables are ratios of squared amplitudes, their dependence on the hadronic
form-factors is partly canceled, which reduces the theoretical uncertainties associated with
the form factors.
Especially $A_{\rm FB}^{}$ is predicted with good precision in the SM to have a zero-crossing
point at \,$q^2\sim4{\rm\,GeV}^2$\,~\cite{Burdman:1995ks}, which makes this asymmetry
very sensitive to the signals of new physics that can shift the location of the point.
In Appendix~\ref{rates} we have written down $F_L^{}$ and $A_{\rm FB}^{}$ in terms
of~$\cal A, C, \ldots, H$.

The amplitude for the contribution of $X$ to \,$B_s\to\ell^+\ell^-$\, can be obtained from
Eq.~(\ref{Hb2sX2sll}) after making the approximation \,$q^2=m_{B_s}^2$\, in the center-of-mass
frame of~$b\bar s$ and neglecting the $\Gamma_X^{}$ part.
Thus, using the matrix elements
\,$\bigl\langle0\bigr|\bar s\gamma^\mu b\bigl|\bar B_s^{}\bigr\rangle =
\bigl\langle0\bigr|\bar s b\bigl|\bar B_s^{}\bigr\rangle = 0$,\,
\,$\bigl\langle0\bigr|\bar s\gamma^\mu\gamma_5^{}b\bigl|\bar B_s^{}(p)\bigr\rangle =
-i f_{B_s}p^\mu$,\,  and
\,$\bigl\langle0\bigr|\bar s\gamma_5^{}b\bigl|\bar B_s^{}\bigr\rangle =
i f_{B_s}m_{B_s}^2/\bigl(m_b^{}+m_s^{}\bigr)$,\,
we arrive at~\cite{Oh:2009fm}
\begin{eqnarray}
{\cal M}_{\bar B_s^{}\to\ell^+\ell^-}^X \,\,=\,\,
-\frac{2i f_{B_s}^{}\, g_{As}^{}\, g_{A\ell}^{}\, m_\ell^{}}{m_X^2}\, \bar\ell\gamma_5^{}\ell ~.
\end{eqnarray}
Since
\,$\bigl\langle0\bigr|\bar s\sigma^{\mu\nu}b\bigl|\bar B_s^{}\bigr\rangle=
\bigl\langle0\bigr|\bar s\sigma^{\mu\nu}\gamma_5^{}b\bigl|\bar B_s^{}\bigr\rangle=0$,\,
the SM yields
\begin{eqnarray}
{\cal M}_{\bar B_s^{}\to\ell^+\ell^-}^{\rm SM} \,\,=\,\,
\frac{-i\alpha G_{\rm F}^{}\lambda_t^{}\,f_{B_s}^{}m_\ell^{}}{\sqrt2\,\pi}\,
C_{10}^{\rm eff}\, \bar\ell\gamma_5^{}\ell ~.
\end{eqnarray}
The sum of the two results in the decay rate
\begin{eqnarray} \label{GB2ll}
\Gamma\bigl(\bar B_s^{}\to\ell^+\ell^-\bigr) \,\,=\,\,
\frac{\alpha^2G_{\rm F}^2\,|\lambda_t^{}|^2f_{B_s}^2\,m_\ell^2}{16\pi^3}
\Biggl|C_{10}^{\rm eff} + \frac{\kappa\, g_{As}^{}\, g_{A\ell}^{}}{m_X^2}\Biggr|^2
\sqrt{m_{B_s}^2-4m_\ell^2} ~.
\end{eqnarray}
Hence \,$B_s^{}\to\ell^+\ell^-$\, does not probe either $g_{Vs}^{}$ or~$g_{V\ell}^{}$.

The $X$ interaction with the lepton $\ell$ as described by ${\cal L}_{\ell X}^{}$ in
Eq.~(\ref{LlX}) affects the anomalous magnetic moment $a_\ell^{}$ of $\ell$ at one loop.
The $X$ contribution to $a_\ell^{}$ can be expressed as~\cite{He:2005we,Leveille:1977rc}
\begin{eqnarray} \label{alX}
a_\ell^X\bigl(m_X^{}\bigr) \,\,=\,\,
\frac{m^2_\ell}{4\pi^2m^2_X}\bigl(g_{V\ell}^2\,f_V^{}(r)+g_{A\ell}^2\,f_A^{}(r)\bigr) ~,
\end{eqnarray}
where \,$r=m^2_\ell/m^2_X$,\,
\begin{eqnarray}
f_V^{}(r) \,\,=\,\, \int^1_0 dx\, \frac{x^2-x^3}{1-x +r x^2} ~, \hspace{5ex}
f_A^{}(r) \,\,=\,\, \int^1_0 dx\, \frac{-4 x+5 x^2-(1+2r)x^3}{1-x +r x^2} ~.
\end{eqnarray}
Since the anomalous magnetic moments of the electron and muon, $a_e^{}$ and $a_\mu^{}$,
have been measured precisely, we need to take their constraints into account.

\section{Numerical analysis\label{numbers}}

\subsection{Constraints}

To obtain the first set of constraints on the contributions of $X$, we employ the data on
the inclusive decay \,$\bar B\to X_s\ell^+\ell^-$.\,
The BaBar and Belle Collaborations have measured its branching ratios $\cal B$ for
different ranges of the squared dilepton invariant mass,~$q^2$.
We take their  results for the low- and high-$q^2$ ranges
\,$1{\rm\,GeV}^2\le q^2\le6{\rm\,GeV}^2$\, and \,$q^2\ge14.4{\rm\,GeV}^2$,\, respectively.
The numbers from BaBar,
\,${\cal B}_{\rm exp}^{\rm low}=(1.8\pm0.7\pm0.5)\times10^{-6}$\, and
\,${\cal B}_{\rm exp}^{\rm high}=
\bigl(5.0\pm2.5_{-0.7}^{+0.8}\bigr)\times10^{-7}$\,~\cite{Aubert:2004it},
and~from Belle,
\,${\cal B}_{\rm exp}^{\rm low}=
\bigl(1.49\pm0.50_{-0.32}^{+0.41}\bigr)\times10^{-6}$\,
and \,${\cal B}_{\rm exp}^{\rm high}=
\bigl(4.2\pm1.2_{-0.7}^{+0.6}\bigr)\times10^{-7}$\,~\cite{Iwasaki:2005sy},
average out to
\begin{eqnarray}
{\cal B}_{\rm exp}^{\rm low}\!\bigl(\bar B\to X_s\ell^+\ell^-\bigr) \,=\,
(1.6\pm0.5)\times10^{-6} ~, \hspace{5ex}
{\cal B}_{\rm exp}^{\rm high}\!\bigl(\bar B\to X_s\ell^+\ell^-\bigr)
\,=\, (4.4\pm 1.2)\times10^{-7} ~. ~~~
\end{eqnarray}
We also need the SM predictions~\cite{Huber:2005ig}
\begin{eqnarray} \label{smB2sll}
{\cal B}_{\rm SM}^{\rm low}\!\bigl(\bar B\to X_s e^+e^-\bigr) =
(1.64\pm0.11)\times10^{-6} ~, &~&
{\cal B}_{\rm SM}^{\rm low}\!\bigl(\bar B\to X_s\mu^+\mu^-\bigr) =
(1.59\pm0.11)\times10^{-6} ~, ~~~ \nonumber \\
{\cal B}_{\rm SM}^{\rm high}\!\bigl(\bar B\to X_s e^+e^-\bigr) =
2.09\times10^{-7}\bigl(1_{-0.30}^{+0.32}\bigr) ~, &&
{\cal B}_{\rm SM}^{\rm high}\!\bigl(\bar B\to X_s\mu^+\mu^-\bigr) =
2.40\times10^{-7}\bigl(1_{-0.26}^{+0.29}\bigr) ~. ~~~~~
\end{eqnarray}
Upon comparing these experimental and SM values, we can require that the $X$ contributions,
from $\Gamma_{b\to s\bar\ell\ell}^X$ in
Eq.~(\ref{Gb2sX}), satisfy
\begin{eqnarray} \label{b2sll_bound}
-5\times10^{-7} \,\le\, {\cal B}_X^{\rm low}\!\bigl(\bar B\to X_s\ell^+\ell^-\bigr) \,\le\,
4\times10^{-7} ~, ~~~~
0 \,\le\, {\cal B}_X^{\rm high}\!\bigl(\bar B\to X_s\ell^+\ell^-\bigr)
\,\le\, 3.5\times10^{-7} ~, ~~~
\end{eqnarray}
where
\begin{eqnarray} \label{xBb2sll}
{\cal B}_X^{}\bigl(\bar B\to X_s\ell^+\ell^-\bigr) \,\,=\,\,
\tau_B^{}\Gamma_{b\to s\bar\ell\ell}^X ~,
\end{eqnarray}
with $\tau_B^{}$ being the $B$ lifetime.
It is clear from Eq.~(\ref{Gb2sX}) that ${\cal B}_X^{}\bigl(\bar B\to X_s\ell^+\ell^-\bigr)$
contains both the $X$-mediated amplitude and its interference with SM~one.

Numerically, we use
\,$\tau_B^{}=\frac{1}{2}\bigl(\tau_{B^+}^{}+\tau_{B^0}^{}\bigr)=1.582$\,ps,\, the average
of the $B^+$ and $B^0$ lifetimes~\cite{pdg},
\,$\alpha=\alpha(m_b^{})=1/133$,\, \,$G_{\rm F}=1.166\times10^{-5}\,{\rm\,GeV}^{-2}$,\,
\,$\lambda_t^{}=V_{ts}^*V_{tb}^{}=-0.0405 + 0.0007i$\,~\cite{ckmfit}, and
the lepton and meson masses from~Ref.~\cite{pdg},\,
as well as the Wilson coefficients \,$C_7^{\rm eff}=-0.304$,\,
\,$C_9^{\rm eff}=4.211+Y(q^2)$,\, and \,$C_{10}^{\rm eff}=-4.103$\, from
Ref.~\cite{Altmannshofer:2008dz}, the expression for the complex function $Y(q^2)$ given therein.
These input parameter values are also used in the rest of the paper.
Since $\Gamma_{b\to s\bar\ell\ell}^{{\rm SM},X}$ in Eq.\,(\ref{Gb2sll}) behave as $m_b^5$ if
$m_b^{}$ gets large, they have sizable uncertainties depending on the choice of~$m_b^{}$ value.
In our numerical treatment of $\Gamma_{b\to s\bar\ell\ell}^X$, we set \,$m_b^{}=4.5$\,GeV,\, as
its application in \,${\cal B}_{\rm SM}^{}=\tau_B^{}\Gamma_{b\to s\bar\ell\ell}^{\rm SM}$\, leads
to \,${\cal B}_{\rm SM}^{\rm low}\!\bigl(\bar B\to X_s\ell^+\ell^-\bigr)=1.70\times10^{-6}$\,
and \,${\cal B}_{\rm SM}^{\rm high}\!\bigl(\bar B\to X_s\ell^+\ell^-\bigr)=1.66\times10^{-7}$,\,
which are compatible with the predicted ranges in Eq.\,(\ref{smB2sll}) from more refined
calculations.

The second set of constraints comes from the data on \,$B\to K^{(*)}\ell^+\ell^-$.\,
The BaBar, Belle, and CDF Collaborations have measured several different observables in these
decays~\cite{Aubert:2008ju,Wei:2009zv,Aaltonen:2011cn}.
We will choose the branching-ratio results provided by Belle and CDF, as they are available for
the specific $q^2$ ranges for which the most recent predictions in the SM with detailed
estimates of the uncertainties are also available.
Their numbers are listed in Tables~\ref{expB2Kll} and~\ref{smB2Kll}, respectively.
In view of the currently sizable experimental and theoretical errors, we will ignore the numerical
effects of $B^+$-$B^0$ and $e$-$\mu$ differences on these processes.
Furthermore, we will take the more precise of each pair of experimental values in
Table~\ref{expB2Kll} and then compare it to the corresponding SM number in Table~\ref{smB2Kll}
in order to estimate the allowed range of the $X$ contribution~${\cal B}_X^{}$.
Accordingly, we can impose the limits
\begin{eqnarray} \label{B2Kll_bound}
-0.7\times10^{-7} \,\,\le\,\,
{\cal B}_X^{}\bigl(\bar B\to\bar K\ell^+\ell^-\bigr)_{q^2\in[1,6]\rm\,GeV^2}
\,\,\le\,\, 0.4\times10^{-7} ~,
\end{eqnarray}
\vspace{-4ex}
\begin{eqnarray*}
-3\times10^{-7} \,\,\le\,\,
{\cal B}_X^{}\bigl(\bar B\to\bar K^*\ell^+\ell^-\bigr)_{q^2\in[1,6]\rm\,GeV^2}
\,\,\le\,\, 0.5\times10^{-7} ~,
\end{eqnarray*}
\vspace{-5ex}
\begin{eqnarray} \label{B2Ksll_bound}
-0.5\times10^{-7} \,\,\le\,\,
{\cal B}_X^{}\bigl(\bar B\to\bar K^*\ell^+\ell^-\bigr)_{q^2\in[14.18,16]\rm\,GeV^2}
\,\,\le\,\, 0.7\times10^{-7} ~,
\end{eqnarray}
\vspace{-5ex}
\begin{eqnarray*}
-0.1\times10^{-7} \,\,\le\,\,
{\cal B}_X^{}\bigl(\bar B\to\bar K^*\ell^+\ell^-\bigr)_{q^2>16\rm\,GeV^2}
\,\,\le\,\, 1.1\times10^{-7} ~,
\end{eqnarray*}
where
\begin{eqnarray} \label{BXB2Kll}
{\cal B}_X^{}\bigl(\bar B\to\bar K^{(*)}\ell^+\ell^-\bigr) \,\,=\,\,
\tau_B^{}\, \Gamma_X^{}\bigl(\bar B\to\bar K^{(*)}\ell^+\ell^-\bigr) ~,
\end{eqnarray}
with  $\Gamma_X^{}\bigl(\bar B\to\bar K^{(*)}\ell^+\ell^-\bigr)$ being the rates of
\,$\bar B\to\bar K^{(*)}\ell^+\ell^-$,\, whose formulas are given in Appendix~\ref{rates},
minus the purely SM part.

The $\bar K^*$ longitudinal polarization fraction $F_L^{}$ and lepton forward-backward asymmetry
$A_{\rm FB}^{}$ in \,$B\to K^*\ell^+\ell^-$\, have also been measured by BaBar, Belle, and~CDF.
Although the BaBar~\cite{Aubert:2008ju} and Belle~\cite{Wei:2009zv} data on
$A_{\rm FB}^{}\bigl(B\to K^*\ell^+\ell^-\bigr)$ exceed the SM expectation, the most recent CDF
measurement~\cite{Aaltonen:2011cn} is consistent with it.
It is expected that more definitive information on this and other observables will be available
with the upcoming results from LHCb in the near future.
Since most of the current data on $F_L^{}$ and $A_{\rm FB}^{}$ still have significant
errors~\cite{Aubert:2008ju,Wei:2009zv,Aaltonen:2011cn}, of order 40\% or greater,
we do not use them in exploring the constraints on $X$.

\begin{table}[t]
\caption{Experimental branching-ratios of \,$B\to K^{(*)}\ell^+\ell^-$\, from
Belle~\cite{Wei:2009zv} and \,$B^{+(0)}\to K^{+(*0)}\mu^+\mu^-$\, from
CDF~\cite{Aaltonen:2011cn}, in units of $10^{-7}$, used to constrain the $X$ contributions,
for different $q^2$ ranges.
The statistical and systematic errors have been combined in quadrature.\label{expB2Kll}}
\medskip
\begin{tabular}{|c|cccc|}
\hline
~$q^2$ $\rm\bigl(GeV^2\bigr)~$ & $~{\cal B}(B\to K\ell^+\ell^-)~$ &
$~{\cal B}(B^+\to K^+\mu^+\mu^-)~$ & $~{\cal B}(B\to K^*\ell^+\ell^-)~$ &
$~{\cal B}(B^0\to K^{*0}\mu^+\mu^-)~$ $\vphantom{\int_|^|}$ \\
\hline\hline
$[1,6]\vphantom{\int^{\int^o}}$ & $1.36_{-0.22}^{+0.24}$ & $1.01\pm 0.27$
& $1.49_{-0.42}^{+0.47}$ & $1.60\pm 0.56$ \\
$~[14.18,16]~\vphantom{\int^|}$ & - & - & $1.05_{-0.27}^{+0.30}$ & $1.51\pm 0.38$ \\
 $>16\vphantom{\int_|^|}$ & - & - & $2.04_{-0.29}^{+0.31}$ & $1.35\pm 0.39$ \\
\hline
\end{tabular}
\end{table}
\begin{table}[t]
\caption{Standard-model predictions for branching-ratios of \,$B\to K^{(*)}\ell^+\ell^-$,\,
in units of $10^{-7}$, for different $q^2$ ranges, from Refs.~\cite{Bobeth:2007dw}.\label{smB2Kll}}
\medskip
\begin{tabular}{|c|cc|}
\hline
~$q^2$ $\rm\bigl(GeV^2\bigr)~$ & $~{\cal B}(B\to K\ell^+\ell^-)~$ &
$~{\cal B}(B\to K^*\ell^+\ell^-)~\vphantom{\int_|^|}$ \\
\hline\hline
$[1,6]\vphantom{\int^{\int^o}}$ & $1.53_{-0.45}^{+0.49}$ & $2.60_{-1.34}^{+1.82}$ \\
$~[14.18,16]~\vphantom{\int^|}$ & - & $1.32_{-0.36}^{+0.43}$ \\
 $>16\vphantom{\int_|^|}$ & - & $1.54_{-0.42}^{+0.48}$ \\
\hline
\end{tabular} \medskip
\end{table}

An additional requirement which the $X$ contributions to \,$B\to K^{(*)}\ell^+\ell^-$\, need to
fulfill is that they do not upset the existing data on the processes
\,$B\to K^{(*)}c\bar c$\, followed by \,$c\bar c\to\ell^+\ell^-$\, in the $q^2$ regions
where the charmonium $(c\bar c)$ resonances lie,
especially if $m_X^2$ also falls within one of these regions.
The relevant charmonia here are $J/\psi$(1S) and $\psi'\equiv\psi$(2S) whose masses are
\,$m_{J/\psi}^{}\simeq3.10$\,GeV\, and \,$m_{\psi'}^{}\simeq3.69$\,GeV\,~\cite{pdg}, respectively.
The constraints follow from comparing the experimental and SM values of the
\,$B\to(J/\psi,\psi')K^{(*)}$, $(J/\psi,\psi')\to\ell^+\ell^-$ branching ratios, which
are not available from the \,$B\to K^{(*)}\ell^+\ell^-$\, measurements,
as events with $q^2$ bins in the neighborhood of $m_{J/\psi,\psi'}^2$ are
vetoed to avoid these long-distance backgrounds mediated by the
resonance~\cite{Aubert:2008ju,Wei:2009zv,Aaltonen:2011cn}.
Experiments on \,$B\to J/\psi K^{(*)}$\, yield
\,${\cal B}(B^+\to J/\psi K^+)=(10.14\pm0.34)\times10^{-4}$,\,
\,${\cal B}(B^0\to J/\psi K^0)=(8.71\pm0.32)\times10^{-4}$,\,
\,${\cal B}(B^+\to J/\psi K^{*+})=(14.3\pm0.8)\times10^{-4}$,\, and
\,${\cal B}(B^0\to J/\psi K^{*0})=(13.3\pm0.6)\times10^{-4}$\,~\cite{pdg},
to be compared to the SM values
\,${\cal B}(B^+\to J/\psi K^+)\!=\!\bigl(9.20^{+6.03}_{-7.99}\bigr)\!\times\!10^{-4}$,\,
\,${\cal B}(B^0\to J/\psi K^0)\!=\!\bigl(8.60^{+5.63}_{-7.47}\bigr)\!\times\!10^{-4}$,\,
\,${\cal B}(B^+\to J/\psi K^{*+})\!=\!\bigl(9.95^{+5.2}_{-7.16}\bigr)\!\times\!10^{-4}$,\, and
\,${\cal B}(B^0\to J/\psi K^{*+})=\bigl(9.30^{+4.86}_{-6.69}\bigr)\times10^{-4}$\,
from Ref.~\cite{Chen:2005ht}.
As for \,$B\to\psi'K$,\, the data are
\,${\cal B}(B^+\to\psi'K^+)=(6.46\pm 0.33)\times10^{-4}$,\,
\,${\cal B}(B^0\to\psi'K^0)=(6.2\pm1.2)\times10^{-4}$,\,
\,${\cal B}(B^+\to\psi'K^{*+})=(6.2\pm0.5)\times10^{-4}$,\, and
\,${\cal B}(B^0\to\psi'K^{*+})=(6.1\pm0.5)\times10^{-4}$\,~\cite{pdg},
whereas the only SM prediction of which we are aware is
\,${\cal B}(B^+\to \psi'K^+)=4.25\times10^{-4}$\, from Ref.~\cite{Gao:2006yu},
without an error estimate.
Since the SM calculations still have considerable uncertainties, after comparing
the experimental and SM numbers for
\,${\cal B}\bigl(B\to(J/\psi,\psi')K^{(*)}\bigr)\,
{\cal B}\bigl((J/\psi,\psi')\to\ell^+\ell^-\bigr)$,\,
with \,${\cal B}(J/\psi\to\ell^+\ell^-)\simeq5.9\%$\, and
\,${\cal B}(\psi'\to\ell^+\ell^-)\simeq0.77\%$\,~\cite{pdg}, we may require
\begin{eqnarray*}
-3\times10^{-5} \,\,\le\,\,
{\cal B}_X^{}\bigl(\bar B\to\bar K\ell^+\ell^-\bigr)_{q^2\in[8.6,10.2]\rm\,GeV^2}
\,\,\le\,\, 5\times10^{-5} ~,
\end{eqnarray*}
\vspace{-5ex}
\begin{eqnarray}  \label{ccbound}
-1\times10^{-5} \,\,\le\,\,
{\cal B}_X^{}\bigl(\bar B\to\bar K^*\ell^+\ell^-\bigr)_{q^2\in[8.6,10.2]\rm\,GeV^2}
\,\,\le\,\, 7\times10^{-5} ~,
\end{eqnarray}
\vspace{-5ex}
\begin{eqnarray*}
-1\times10^{-6} \,\,\le\,\,
{\cal B}_X^{}\bigl(\bar B\to\bar K^{(*)}\ell^+\ell^-\bigr)_{q^2\in[12.8,14.2]\rm\,GeV^2}
\,\,\le\,\, 4\times10^{-6} ~,
\end{eqnarray*}
where the chosen $q^2$ bins are similar to those for the charmonium backgrounds
in the experiments.

To evaluate ${\cal B}_X^{}\bigl(\bar B\to\bar K^{(*)}\ell^+\ell^-\bigr)$,
we employ the \,$B\to K^{(*)}$ form factors from light-cone sum rules~\cite{Ball:2004ye,Ball:2004rg}.
We have collected the formulas of the form factors in Appendix~\ref{ff}.
In our search for the allowed parameter space of the $X$ couplings, we use the lower bounds
of the form factors in order to get the most space.

\begin{figure}[b]
\includegraphics[width=177mm]{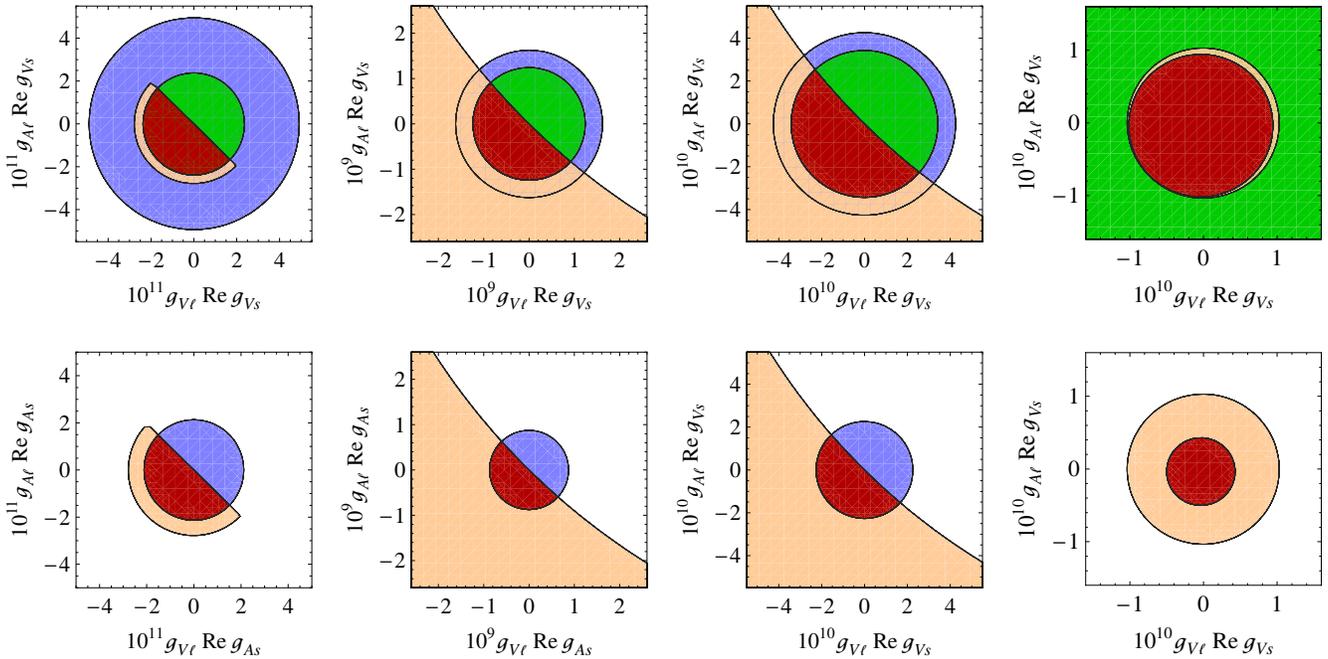}
\caption{Regions of \,$\bigl(g_{V\ell}^{},g_{A\ell}^{}\bigr){\rm Re}\,g_{Vs}^{}$\,
for \,${\rm Im}\,g_{Vs}^{}=g_{As}^{}=0$\, (top plots) and of
\,$\bigl(g_{V\ell}^{},g_{A\ell}^{}\bigr){\rm Re}\,g_{As}^{}$\,
for \,${\rm Im}\,g_{As}^{}=g_{Vs}^{}=0$\, (bottom plots) satisfying constraints from
\,$\bar B\to X_s^{}\ell^+\ell^-$ (orange, lightly shaded),
$\bar B\to\bar K\ell^+\ell^-$ (green, medium shaded),
$\bar B\to\bar K^*\ell^+\ell^-$ (blue, heavily shaded), and all of them (dark red).
From left to right, the plots correspond to \,$m_X^{}=2$, 3, 3.7,
and 4~GeV.\label{realgVgA}}
\end{figure}

After scanning the relevant parameters, we have found that there is $X$ parameter space
available which satisfies the constraints in Eqs.~(\ref{b2sll_bound}), (\ref{B2Kll_bound}),
(\ref{B2Ksll_bound}), and~(\ref{ccbound}).
The size of the allowed parameter space can vary widely, depending on the mass and
total width of~$X$.
To illustrate this, we take the $X$-mass values \,$m_X^{}=2$, 3, 3.7, and~4~GeV.\,
Since the couplings of $X$ to other fermions are not specified in our approach,
its total width is unknown.
For definiteness we choose \,$\Gamma_X^{}=0.1$\,MeV\, in the four cases, but will comment
later on other choices.
Since in the decay amplitudes the couplings $g_{Vs}^{}$ and $g_{As}^{}$ always occur each
multiplied by $g_{V\ell}^{}$ or $g_{A\ell}^{}$, we show in Fig.~\ref{realgVgA} the allowed
regions of their products in the cases of $g_{Vs}^{}$ being real with \,$g_{As}^{}=0$ (top plots)
and $g_{As}^{}$ being real with \,$g_{Vs}^{}=0$ (bottom plots).
In the latter, \,$g_{Vs}^{}=0$,\, cases, the restrictions from \,$B\to K\ell^+\ell^-$\,
are absent because $g_{As}^{}$ is not present in its amplitude.
In the two rightmost plots, for \,$m_X^{}=4$\,GeV,\, the blue areas are invisible because
they coincide exactly with the red ones.
One could also get allowed parameter space for both $g_{Vs,As}^{}$ being nonzero
and one or both of them complex.
We will have some examples with the complex couplings shortly.

\begin{figure}[b] \vspace{2ex}
\includegraphics[height=45mm,width=47mm]{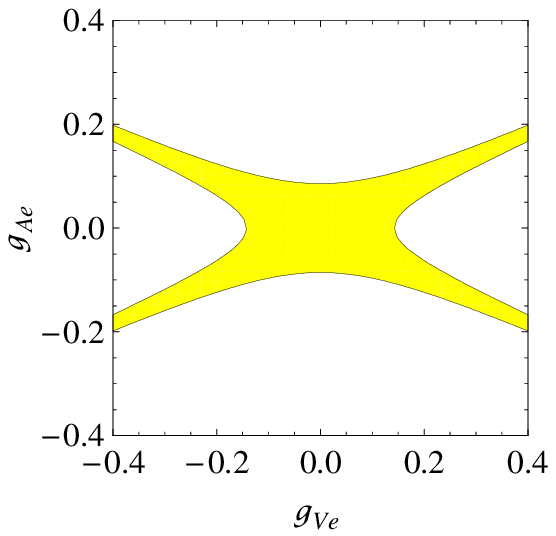} \, \, \,
\includegraphics[height=45mm,width=50mm]{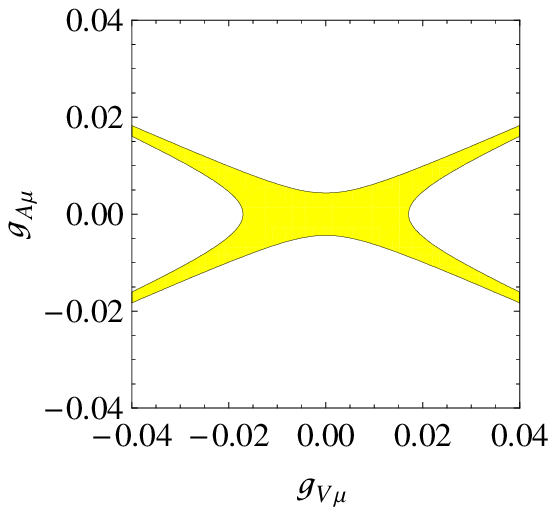} \vspace*{-1ex}
\caption{Parameter space of $\bigl(g_{Ve}^{},g_{Ae}^{}\bigr)$ and
$\bigl(g_{V\mu}^{},g_{A\mu}^{}\bigr)$ for \,$m_X^{}=3$\,GeV\, subject to constraints from
the anomalous magnetic moments of the electron and muon, respectively.\label{alX_bounds}}
\end{figure}

If one deals with $g_{Vs,As}^{}$ separately from $g_{V\ell,A\ell}^{}$, instead of their products,
it will be necessary to take into account constraints on the latter from the anomalous magnetic
moments of the electron and muon, $a_e^{}$ and $a_\mu^{}$.
From their measurements, we have
\,$a_e^{\rm exp}=(115965218073\pm 28)\times10^{-14}$\, and
\,$a_\mu^{\rm exp}=(11659209\pm 6)\times 10^{-10}$\,~\cite{pdg}.
The SM prediction for $a_e^{}$ agrees with its measurement,
\,$a_e^{\rm exp}-a_e^{\rm SM}=(-206\pm 770)\times10^{-14}$,\, but the SM prediction for
$a_\mu^{}$ presently differs by $3.2\sigma$ from its experimental value,
\,$a_\mu^{\rm exp}-a_\mu^{\rm SM}=(29\pm9)\times10^{-10}$\,~\cite{Jegerlehner:2009ry}.
Consequently, since the $g_{V\ell}^{}$ and $g_{A\ell}^{}$ terms in $a_\ell^X$ are opposite
in sign, we may impose
\begin{eqnarray} \label{a_req}
-9\times10^{-12} \,\,\le\,\, a_e^X \,\,\le\,\, 5\times10^{-12} ~, \hspace{5ex}
-1\times10^{-9} \,\,\le\,\, a_\mu^X \,\,\le\,\, 3\times10^{-9} ~.
\end{eqnarray}
Applying \,$m_X^{}=2$, 3, 3.7, 4~GeV\, in the $X$ contributions $a_\ell^X$ in Eq.~(\ref{alX}) yields
\begin{eqnarray} \label{al}
a_e^X(2{\rm\,GeV}) \,=\, \bigl(5.5\,g_{Ve}^2 - 27.6\,g_{Ae}^2\bigr)\times10^{-10} ~, &~~~&
a_\mu^X(2{\rm\,GeV}) \,=\, \bigl(22.8\,g_{V\mu}^2 - 117\,g_{A\mu}^2\bigr)\times10^{-6} ~,
\nonumber \\
a_e^X(3{\rm\,GeV}) \,=\, \bigl(2.4\,g_{Ve}^2 - 12.2\,g_{Ae}^2\bigr)\times10^{-10} ~, &~~~&
a_\mu^X(3{\rm\,GeV}) \,=\, \bigl(10.3\,g_{V\mu}^2 - 52.2\,g_{A\mu}^2\bigr)\times10^{-6} ~,
\nonumber \\
a_e^X(3.7{\rm\,GeV}) \,=\, \bigl(1.6\,g_{Ve}^2 - 8.1\,g_{Ae}^2\bigr)\times10^{-10} ~, &~~~&
a_\mu^X(3.7{\rm\,GeV}) \,=\, \bigl(6.8\,g_{V\mu}^2 - 34.3\,g_{A\mu}^2\bigr)\times10^{-6} ~, ~~~~
\nonumber \\
a_e^X(4{\rm\,GeV}) \,=\, \bigl(1.4\,g_{Ve}^2 - 6.9\,g_{Ae}^2\bigr)\times10^{-10} ~, &~~~&
a_\mu^X(4{\rm\,GeV}) \,=\, \bigl(5.8\,g_{V\mu}^2 - 29.4\,g_{A\mu}^2\bigr)\times10^{-6} ~. ~~
\end{eqnarray}
To illustrate the parameter space of $g_{V\ell}^{}$ and $g_{A\ell}^{}$ subject to the bounds
in Eq.~(\ref{a_req}), we display in Fig.~\ref{alX_bounds} the \,$m_X^{}=3$\,GeV\, case.
One can conclude from Eq.~(\ref{al}) and these plots that for each value of $m_X^{}$ the allowed
$\bigl(g_{V\ell}^{},g_{A\ell}^{}\bigr)$ region for \,$\ell=e$\, is much larger than that for
\,$\ell=\mu$,\, although the reverse is true for the imposed limits on $a_e^X$ and~$a_\mu^X$
in Eq.~(\ref{a_req}).

\begin{figure}[t]
\includegraphics[width=177mm]{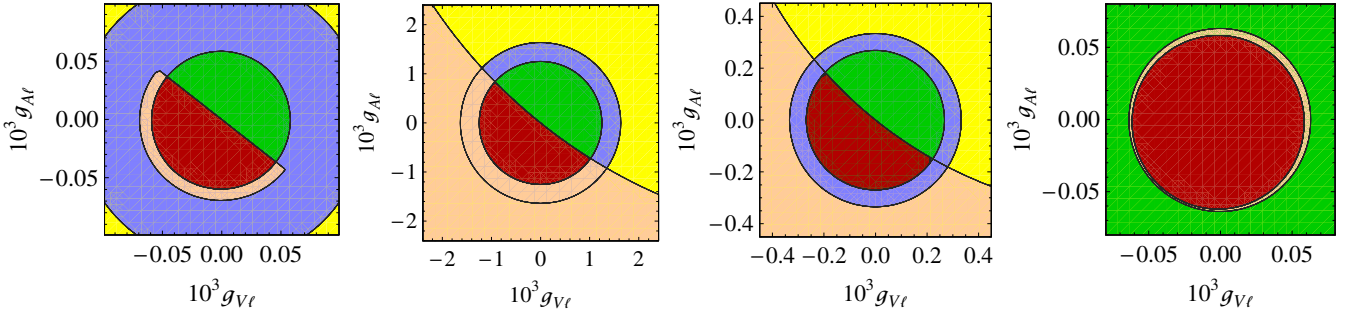}
\caption{Allowed ranges of $g_{V\ell}^{}$ and $g_{A\ell}^{}$ for (from left to right)
\,$m_X^{}=2$, 3, 3.7, and 4~GeV\, with \,$g_{As}^{}=0$\, and $g_{Vs}^{}$ values given
in the text, subject to constraints from  \,$B\to X_s^{}\ell^+\ell^-$ (orange, lightly shaded),
$B\to K\ell^+\ell^-$ (green, medium shaded), $B\to K^*\ell^+\ell^-$ (blue, heavily shaded),
$a_\ell^{}$ (yellow, very lightly shaded), and all of them~(dark red).\label{complexgVnogA}}
\end{figure}

We now explore the allowed $g_{V\ell,A\ell}^{}$ ranges for $g_{Vs,As}^{}$ values that can
lead to predictions compatible, at the one-sigma level, with the D0 anomalous measurement
of the like-sign dimuon charge asymmetry in semileptonic $b$-hadron decays~\cite{Abazov:2010hv},
as we proposed in Ref.~\cite{Oh:2010vc}.
This implies that at least one of $g_{Vs,As}^{}$ has to be complex.
For \,$m_X^{}=2$, 3, 3.7, and 4~GeV,\, using the results of Ref.~\cite{Oh:2010vc}, we obtain
\,$g_{Vs}^{}=(3.0-2.7i)\times10^{-7}$, $(7.5-6.5i)\times10^{-7}$,  $(9.5-8.5i)\times10^{-7}$,
and $(12-11i)\times10^{-7}$,\, respectively, as possible values in the \,$g_{As}^{}=0$\, case.
In~Fig.~\ref{complexgVnogA} we show the corresponding regions consistent with the various
constraints described above.
The areas displayed in the four plots have all turned out to be well within the bounds from
both $a_e^{}$ and $a_\mu^{}$ (yellow, very lightly shaded, if not covered by the other bounds).
In the fourth plot, for \,$m_X^{}=4$\,GeV,\, the blue area coincides exactly with,
and hence is completely covered by, the red one.
If \,$g_{Vs}^{}=0$,\, one could also find the allowed parameter space, in which case
the restrictions from \,$B\to K\ell^+\ell^-$ (blue areas) are again absent and
the regions consistent with all the other constraints are generally smaller than in
the cases with \,$g_{As}^{}=0$.\,

\subsection{Predictions}

From the values of the $X$ couplings $g_{Vs,As}^{}$ and $g_{V\ell,A\ell}^{}$ allowed by
the various limits above, we can assess the effects that $X$ might have on observables which
will likely be studied experimentally in the near future.
With improved precision compared to the existing data, the upcoming measurements will test
the existence of $X$ more stringently or place stricter constraints on its couplings.
The observables we discuss here are the differential branching ratios of
\,$\bar B\to\bar K^{(*)}\ell^+\ell^-$,\, the $\bar K^*$ longitudinal polarization fraction
$F_L^{}$ in \,$\bar B\to\bar K^*\ell^+\ell^-$,\, and the lepton forward-backward asymmetry
$A_{\rm FB}^{}$ in the latter decay, as well as the branching ratios of some rare $B_s^{}$ decays.

\begin{figure}[t]
\includegraphics[height=61mm,width=85mm]{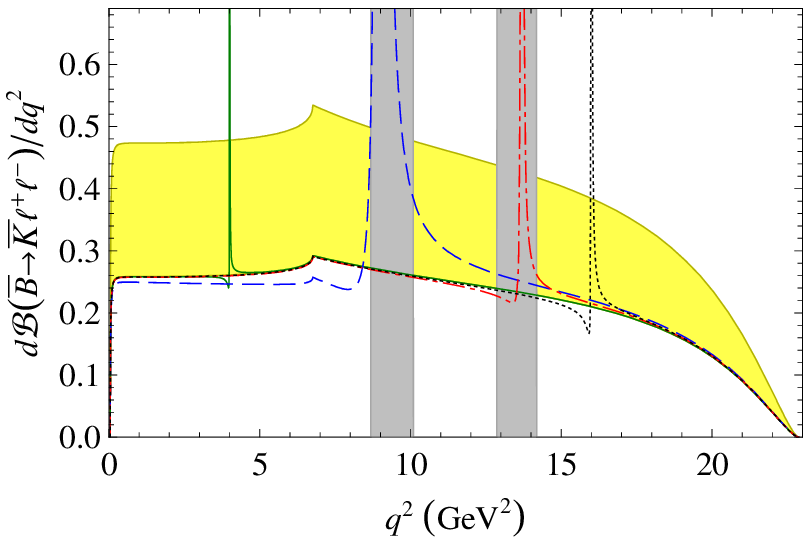} \, \,
\includegraphics[height=61mm,width=85mm]{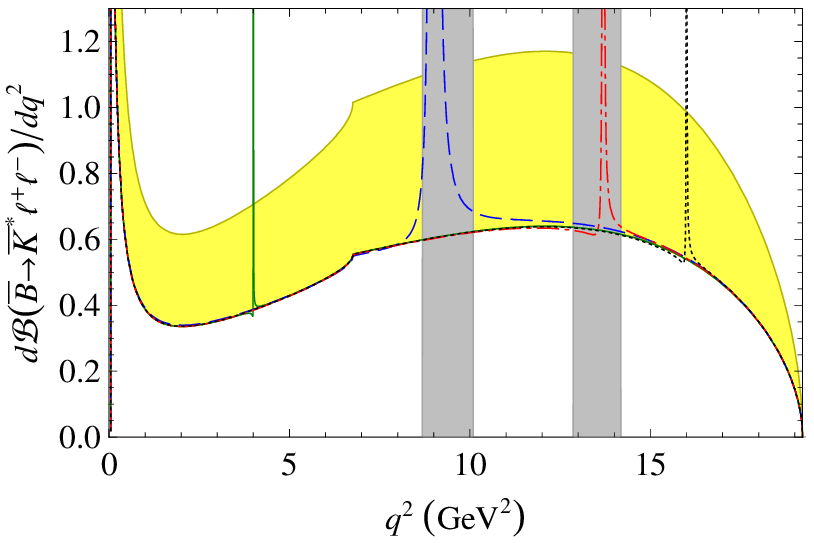} \vspace*{-1ex}
\caption{Differential branching ratios of \,$\bar B\to\bar K\ell^+\ell^-$ (left plot)  and
\,$\bar B\to\bar K^*\ell^+\ell^-$ (right plot) as functions of the squared dilepton-mass
in the SM (yellow curved bands) and its combination with the $X$ contribution for
\,$m_X^{}=2$ (green solid curves), 3 (blue dashed curves),
3.7 (red dot-dashed curves), and 4 (black dotted curves) GeV,\,
with the $g_{Vs,As}^{}$ and $g_{V\ell,A\ell}^{}$ values in Eq.~(\ref{gg}).\label{dB/ds}}
\end{figure}

To illustrate how the $X$ contributions may modify the SM predictions, we adopt for definiteness
some of the larger values of the couplings from the top plots in Fig.~\ref{realgVgA}.
Thus we have
\begin{eqnarray} \label{gg}
\bigl(g_{V\ell}^{},\,g_{A\ell}^{}\bigr)g_{Vs}^{} \,\,= \left\{
\begin{array}{ll}     \displaystyle
(1.0,\,-2.0)\times10^{-11} & ~~~ \mbox{for \, $m_X^{}=2$ GeV}  \\
(-1.0,\,0.5)\times10^{-9}  & ~~~ \mbox{for \, $m_X^{}=3$ GeV}  \\
(1.0,\,-2.0)\times10^{-10} & ~~~ \mbox{for \, $m_X^{}=3.7$ GeV}  \\
(-9.0,\,3.0)\times10^{-11} & ~~~ \mbox{for \, $m_X^{}=4$ GeV}
\end{array}   \right.
\end{eqnarray}
with  \,$g_{As}^{}=0$.\,
These numbers translate into the differential branching ratios of
\,$\bar B\to\bar K^{(*)}\ell^+\ell^-$\, in Fig.~\ref{dB/ds}, where the curved (yellow) bands
correspond to the SM expectation including the $\pm15\%$ uncertainties in the form factors and
for the curves corresponding to the combined SM and $X$ contribution we have used the lower
limits of the form factors.
The two vertical (gray) bands mark $q^2$ ranges in which decay events are vetoed in
the \,$\bar B\to\bar K^{(*)}\ell^+\ell^-$\, experiments to reject backgrounds from the charmonium
resonances~\cite{Aubert:2008ju,Wei:2009zv,Aaltonen:2011cn}.

From Fig.~\ref{dB/ds} one can see that the impact of $X$ can lead to mild changes in
the SM branching ratios.
In particular, the differential branching ratios may have spikes, indicating the $X$ presence
at \,$q^2=m_X^2$,\, possibly accompanied by small dips and rises on opposite sides of the spikes
arising from the enhanced interference near the resonant point between the SM and $X$ amplitudes.
To observe these features would require not only high precision in measuring the branching
ratios in the $q^2$ bins, but also sufficiently small bin sizes.
Both are not yet available in the existing data from BaBar, Belle,
and~CDF~\cite{Aubert:2008ju,Wei:2009zv,Aaltonen:2011cn}.

\begin{figure}[b] \vspace*{2ex}
\includegraphics[height=62mm,width=85mm]{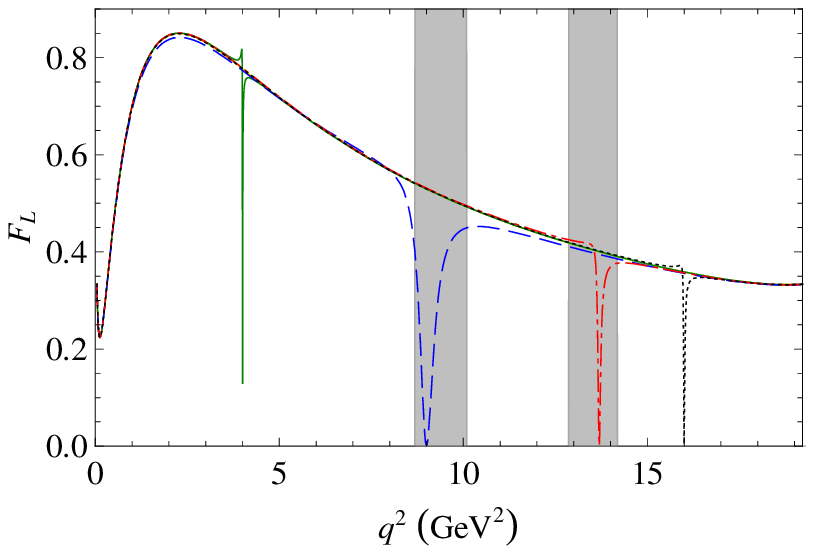} \, \,
\includegraphics[height=62mm,width=85mm]{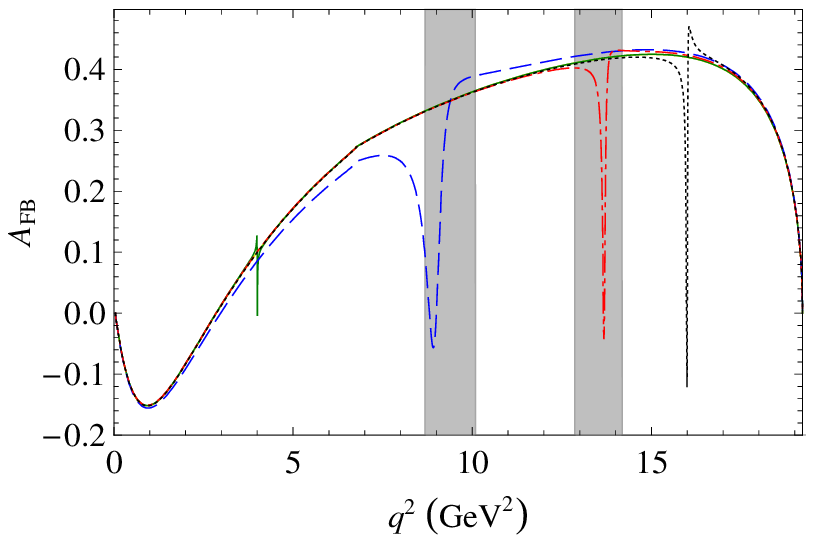} \vspace*{-1ex}
\caption{Plots of $\bar K^*$ longitudinal polarization fraction (left) and
lepton forward-backward asymmetry (right) for \,$\bar B\to\bar K^*\ell^+\ell^-$\,
in the SM (solid curves) and its combination with the $X$ contribution for
\,$m_X^{}=2$ (green solid curves), 3 (blue dashed curves),
3.7 (red dot-dashed curves), and 4 (black dotted curves) GeV,\,
with the $g_{Vs,As}^{}$ and $g_{V\ell,A\ell}^{}$ values in Eq.~(\ref{gg}).\label{flafb}}
\end{figure}

With the couplings in Eq.~(\ref{gg}) as before, we graph in Fig.~\ref{flafb} the $\bar K^*$
longitudinal polarization fraction $F_L^{}$ and lepton forward-backward asymmetry
$A_{\rm FB}^{}$ for \,$\bar B\to\bar K^*\ell^+\ell^-$.\,
These plots suggest that, since the SM uncertainties are much reduced for these observables,
the contributions of $X$ to them, especially the latter, can produce modifications to the SM
predictions that are likely to be more detectable than in the branching ratios.
Thus the small spikes and deep narrow dips around \,$q^2=m_X^2$,\, corresponding to the dips
and spikes in the differential branching ratios, could be very revealing.
However, as in the branching-ratio case, it would be necessary to have high precision in
the $F_L^{}$ and $A_{\rm FB}^{}$ measurements, as well as small sizes of the $q^2$ bins.
This applies particularly to, for instance, the determination of the location of the $A_{\rm FB}^{}$
zero-crossing point which may only be slightly affected by the presence of $X$, as
the right plot in Fig.~\ref{flafb} indicates.
In contrast, the available experimental results on $F_L^{}$ and $A_{\rm FB}^{}$ still have
substantial relative errors, more so than the branching
ratios~\cite{Aubert:2008ju,Wei:2009zv,Aaltonen:2011cn}.

At this point we should comment that the effects of $X$ as depicted in Figs.~\ref{dB/ds}
and~\ref{flafb} correspond to the choice for its total width \,$\Gamma_X^{}=100$\,keV\,
which we made earlier.
If a larger (smaller) value of $\Gamma_X^{}$ were assumed, the parameter space of $X$
consistent with the constraints we discussed would be larger (smaller) also, and
the signals of $X$ would become more (less) detectable.
In any case, the availability of the allowed parameter space should encourage experimental
searches for $X$ in future $b$-hadron experiments.

Finally, we turn to a few rare $B_s^{}$ decays, the first one being
\,$\bar B_s^{}\to\phi\ell^+\ell^-$,\, which also proceeds from \,$b\to s\ell^+\ell^-$.\,
The expressions for its amplitude and decay rate can then be obtained from those for
\,$\bar B\to\bar K^*\ell^+\ell^-$,\, with appropriate changes.
Using the \,$\bar B_s^{}\to\phi$\, form-factors listed in Appendix~\ref{ff}, we
estimate the SM branching ratio to be
\,$1.0\times10^{-6}\le{\cal B}_{\rm SM}^{}\bigl(\bar B_s^{}\to\phi\ell^+\ell^-\bigr)\le
1.8\times10^{-6}$,\,
after excluding \,$8.68{\rm\,GeV}^2\le q^2\le10.09{\rm\,GeV}^2$\, and
\,$12.86{\rm\,GeV}^2\le q^2\le14.18{\rm\,GeV}^2$\, from the $q^2$ integration,
as in the experimental studies.
Very recently CDF has reported the first observation of this decay, with
\,${\cal B}_{\rm exp}^{}\bigl(\bar B_s^{}\to\phi\mu^+\mu^-\bigr)=
(1.44\pm0.33\pm0.46)\times10^{-6}$\,~\cite{Aaltonen:2011cn},
which is compatible with the SM value.
In view of the still-large theoretical and experimental errors, there is room in this
decay for the $X$ contribution, which can expectedly be tested with improved precision
at~LHCb.
We further expect that the $\phi$ longitudinal polarization fraction and lepton
forward-backward asymmetry in this decay are sensitive tools to detect the signals of $X$,
much like their counterparts in~\,$\bar B\to\bar K^*\ell^+\ell^-$.\,

As mentioned earlier, the decays \,$B_s^{}\to(\eta,\eta')\ell^+\ell^-$\, are also relevant in the search
for $X$, but they are yet to be observed.
They may be detected in the near future, as the SM values of their branching ratios are estimated
to be only several times smaller than those of \,$B_s^{}\to\phi\ell^+\ell^-$\,~\cite{Geng:2003su}.
Since the amplitudes for \,$B_s^{}\to(\eta,\eta')\ell^+\ell^-$\, are independent of
the coupling~$g_{As}^{}$, as in the case of \,$\bar B\to\bar K\ell^+\ell^-$,\, we expect that
qualitatively the effects of $X$ on the former would be roughly similar to that on the latter.

Additional decays of interest which are not yet observed are \,$B_s^{}\to\ell^+\ell^-$.\,
Their amplitudes do not involve~$g_{Vs}^{}$, in contrast to \,$B_s^{}\to(\eta,\eta')\ell^+\ell^-$.\,
The experimental upper-bounds on their branching ratios are available~\cite{pdg},
\begin{eqnarray}
{\cal B}_{\rm exp}^{}\bigl(B_s^{}\to e^+e^-\bigr) \,\,<\,\, 2.8\times10^{-7}  ~, \hspace{5ex}
{\cal B}_{\rm exp}^{}\bigl(B_s^{}\to\mu^+\mu^-\bigr) \,\,<\,\, 3.2\times10^{-8} ~.
\end{eqnarray}
From Eq.~(\ref{GB2ll}), we find the SM values
\begin{eqnarray}
{\cal B}_{\rm SM}^{}\bigl(B_s^{}\to e^+e^-\bigr) \,\,\simeq\,\, 7.5\times10^{-14}  ~, \hspace{5ex}
{\cal B}_{\rm SM}^{}\bigl(B_s^{}\to\mu^+\mu^-\bigr) \,\,\simeq\,\, 3.2\times10^{-9}
\end{eqnarray}
using the central value of \,$f_{B_s}=240\pm30$\,MeV\,~\cite{Onogi:2006km}.
If we include the contribution of $X$ in Eq.~(\ref{GB2ll}), we have, for instance,
\,$-9\times10^{-10}\lesssim g_{As}^{}g_{A\ell}^{}\lesssim6\times10^{-10}$\,
from the second of the bottom plots in~Fig.~\ref{realgVgA}, which translates into
\begin{eqnarray}
6.6\times10^{-14} &\lesssim& {\cal B}\bigl(B_s^{}\to e^+e^-\bigr) \,\,\lesssim\,\,
8.1\times10^{-14}  ~, \nonumber \\
2.8\times10^{-9} &\lesssim& {\cal B}\bigl(B_s^{}\to\mu^+\mu^-\bigr) \,\,\lesssim\,\,
3.5\times10^{-9}  ~.
\end{eqnarray}
Hence the $X$ contributions are easily accommodated by the present experimental limits
and can produce moderate modifications to the SM predictions.
Once \,$B_s^{}\to\mu^+\mu^-$\, has been seen, it can provide important restrictions on
the combination \,$g_{As}^{}g_{A\mu}^{}$\, if its branching ratio turns out
be consistent with the SM expectation.

\section{Conclusions\label{concl}}

We have extended our proposal in an earlier paper that a nonstandard spin-1 particle lighter
than the $b$~quark with flavor-changing couplings to $b$ and $s$ quarks can offer a viable
explanation for the recent anomalous measurement by D0 of the like-sign dimuon charge asymmetry
in semileptonic $b$-hadron decays.
Specifically we have considered the possibility that the new particle also couples to the light
charged leptons \,$\ell=e,\mu$\, and thus contributes to rare \,$b\to s$\, transitions involving
the leptons.
After exploring experimental constraints on its couplings from currently available
experimental data on the inclusive \,$B\to X_s\ell^+\ell^-$\, and exclusive
\,$B\to K^{(*)}\ell^+\ell^-$\, decays, as well as the charmonium contributions to
the latter and the anomalous magnetic moments of the leptons,
we have shown that there is parameter space of the particle that is consistent with the data,
including the D0 result.
This can serve to help motivate dedicated searches for $X$ in $b$-hadron experiments in the future.
We have found that the differential branching ratios of these decays, the $K^*$ longitudinal
polarization in \,$B\to K^*\ell^+\ell^-$,\, and its lepton forward-backward
asymmetry, especially the latter two, are observables that are potentially sensitive to
the signals of $X$.
Furthermore, rare decays of the $B_s$ meson, such as \,$B_s\to(\phi,\eta,\eta')\ell^+\ell^-$\, and
\,$B_s\to\ell^+\ell^-$,\, can provide additional observables.
We expect that the upcoming measurements of rare $B$ and $B_s^{}$ processes with greatly increased
precision at LHCb and next-generation $B$ factories can probe the existence of the particle,
or its couplings, stringently.

\acknowledgments \vspace{-2ex}
This work was supported in part by the National Science Council of R.O.C. under grants
numbers NSC-99-2811-M-001-038 and NSC-99-2811-M-008-019.

\appendix

\section{Form factors\label{ff}}

To obtain the \,$\bar B\to\bar K\ell^+\ell^-$\, amplitude, we use the matrix elements
\begin{eqnarray*}
\bigl\langle\bar K\bigr|\bar s\gamma^\mu b\bigl|\bar B\bigr\rangle \,\,=\,\,
\frac{m_B^2-m_K^2}{q^2}\, q^\mu\, F_0^{} +
\biggl(p_B^\mu+p_K^\mu-\frac{m_B^2-m_K^2}{q^2}\,q^\mu\biggr) F_1^{} ~,
\end{eqnarray*}
\vspace{-4ex}
\begin{eqnarray*}
\bigl\langle\bar K\bigr|\bar s b\bigl|\bar B\bigr\rangle \,\,=\,\,
\frac{m_B^2-m_K^2}{m_b^{}-m_s^{}}\, F_0^{} ~,
\end{eqnarray*}
\vspace{-4ex}
\begin{eqnarray*}
q_\nu^{}\bigl\langle\bar K\bigr|\bar s\sigma^{\mu\nu}b\bigl|\bar B\bigr\rangle \,\,=\,\,
\frac{q^2\,(p_B^{}+p_K^{})^\mu-\bigl(m_B^2-m_K^2\bigr)q^\mu}{m_B^{}+m_K^{}}\,iF_T^{} ~,
\end{eqnarray*}
\vspace{-4ex}
\begin{eqnarray}
\langle\bar K|\bar s\gamma^\mu\gamma_5^{}b|\bar B\rangle \,\,=\,\,
\langle\bar K|\bar s\gamma_5^{}b|\bar B\rangle \,\,=\,\,
q^\nu\bigl\langle\bar K\bigr|\bar s\sigma_{\mu\nu}^{}\gamma_5^{}b\bigl|\bar B\bigr\rangle
\,\,=\,\, 0 ~,
\end{eqnarray}
where $p_B^{}$ $(p_K^{})$ is the $\bar B$ $\bigl(\bar K\bigr)$ momentum,
\,$q=p_B^{}-p_K^{}$,\, and the form factors $F_{0,1,T}^{}$ each depend on~$q^2$.
The matrix elements relevant to \,$\bar B\to\bar K^*\ell^+\ell^-$,\, are
\begin{eqnarray*}
\bigl\langle\bar K^*\bigr|\bar s\gamma_\mu^{}b\bigl|\bar B\bigr\rangle &=&
\frac{2 V^{}}{m_B^{}+m_{K^*}^{}}\, \epsilon_{\mu\nu\sigma\tau}^{}\,
\varepsilon^{*\nu} p_B^\sigma\, p_{K^*}^\tau ~,
\end{eqnarray*}
\vspace{-4ex}
\begin{eqnarray*}
\bigl\langle\bar K^*\bigr|\bar s\gamma^\mu\gamma_5^{}b\bigl|\bar B\bigr\rangle  &=&
2 i A_0^{}\, m_{K^*}^{}\, \frac{\varepsilon^*\!\cdot\!q}{q^2}\, q^\mu
+ i A_1^{}\bigl( m_B^{}+m_{K^*}^{} \bigr) \biggl(
\varepsilon^{*\mu}-\frac{\varepsilon^*\!\cdot\!q}{q^2}\, q^\mu \biggr)
\\ && - \,\,
\frac{i A_2^{}\,\varepsilon^*\!\cdot\!q}{m_B^{}+m_{K^*}^{}}
\biggl( p_B^\mu + p_{K^*}^\mu - \frac{m_B^2-m_{K^*}^2}{q^2}\, q^\mu \biggr)_{\vphantom{|}} ~,
\end{eqnarray*}
\vspace{-4ex}
\begin{eqnarray*}
\bigl\langle\bar K^*\bigr|\bar s b\bigl|\bar B\bigr\rangle \,\,=\,\, 0 ~, \hspace{5ex}
\bigl\langle\bar K^*\bigl(p_{K^*}^{}\bigr)\bigr|\bar s\gamma_5^{}b
\bigl|\bar B\bigl(p_B^{}\bigr)\bigr\rangle \,\,=\,\,
\frac{-2i A_0^{}\,m_{K^*}^{}\,\varepsilon^{*}\!\cdot\!q}{m_b^{}+m_s^{}} ~,
\end{eqnarray*}
\vspace{-4ex}
\begin{eqnarray*}
q^\nu\bigl\langle\bar K^*\bigr|\bar s\sigma_{\mu\nu}^{}b\bigl|\bar B\bigr\rangle \,\,=\,\,
2i T_1^{}\epsilon_{\mu\nu\sigma\tau}^{}\,
\varepsilon^{*\nu} p_B^\sigma\, p_{K^*}^\tau ~,
\end{eqnarray*}
\vspace{-4ex}
\begin{eqnarray}
q^\nu\bigl\langle\bar K^*\bigr|\bar s\sigma_{\mu\nu}^{}\gamma_5^{}b
\bigl|\bar B\bigr\rangle &=&
T_2^{}\Bigl[ \bigl(m_B^2-m_{K^*}^2\bigr)\varepsilon_\mu^{*} -
\bigl(p_B^{}+p_{K^*}^{}\bigr)_{\!\mu} \varepsilon^{*}\!\cdot\!q \Bigr]
\nonumber \\ && +\;
T_3^{} \Biggl[ q_\mu^{} - \frac{(p_B^{}+p_{K^*}^{})_\mu^{} q^2}{m_B^2-m_{K^*}^2} \Biggr]
\varepsilon^{*}\!\cdot\!q ~,
\end{eqnarray}
where $p_{K^*}^{}$ is the $\bar K^*$ momentum, \,$q=p_B^{}-p_{K^*}^{}$,\, and
the form factors $V^{}$, $A_{0,1,2}^{}$, and $T_{1,2,3}^{}$ are all functions of~$q^2$.

\begin{table}[b]
\caption{Parameters for \,$B\to K^{(*)}$\, form factors~\cite{Ball:2004ye,Ball:2004rg}.
The $m_{\rm fit}^2$ and $m_R^{}$ entries are in units of $\rm GeV^2$ and GeV, respectively.}
\medskip
\begin{tabular}{|c|cccccccccc|}
\hline
$\vphantom{\int_|^|}$ & $F_0^{}$ & $F_1^{}$ & $F_T^{}$ & $A_0^{}$ & $A_1^{}$ & $A_2^{}$ & $V$ &
$T_1^{}$ & $T_2^{}$ & $\tilde T_3^{}$ \\
\hline\hline
$~r_1^{\vphantom{|}}~$ &
   --   &  0.162  &  0.161  &  ~1.364    &   --   & ~~$-0.084$~~ & ~0.923 &  ~0.823 &  --   & ~$-0.036$~ \\
$r_{2_{\vphantom{|}}}^{\vphantom{|}}$ &
~0.330~ & ~~0.173~~ & ~0.198~ & ~~$-0.990$~~ & ~0.292~ & $~~0.342$~ & ~$-0.511$~ & ~~$-0.491$~~ & ~~$0.333$~~ & ~$~0.368$~ \\
~$m_{\rm fit_{\vphantom{|}}}^2$~ &
                  37.46 & --      & --     & ~36.78 &  40.38 & ~~52.00 &  ~49.40 &  ~46.31 & 41.41 &  ~48.10 \\
~$m_{R_{\vphantom{|}}}^{}$~ &
--   &   5.41 &   5.41 &   ~~5.28 &  --  &  --    & ~~5.32 & ~~5.32 &  --   &  --    \\
\hline
\end{tabular} \label{ffparameters}
\end{table}

In numerical calculations, we adopt the form-factor results of Refs.~\cite{Ball:2004ye,Ball:2004rg}
from light-cone sum rules.
The form factors are parametrized as
\begin{eqnarray}
\textsf{\textsl F}(s) \,=\, \frac{r_1^{}}{1-s/m_R^2} + \frac{r_2^{}}{\bigl(1-s/m_R^2\bigr)^2}
\hspace{5ex} \mbox{for \, $F_1^{}$\,, \,$F_T^{}$} ~,
\end{eqnarray}
\vspace{-4ex}
\begin{eqnarray}
\textsf{\textsl F}(s) \,\,=\,\, \frac{r_2^{}}{1-s/m_{\rm fit}^2}
\hspace{5ex} \mbox{for \, $F_0^{}$\,, \,$A_1^{}$\,, \,$T_2^{}$} ~,
\end{eqnarray}
\vspace{-4ex}
\begin{eqnarray}
\textsf{\textsl F}(s) \,\,=\,\, \frac{r_1^{}}{1-s/m_R^2} \,+\, \frac{r_2^{}}{1-s/m_{\rm fit}^2}
\hspace{5ex} \mbox{for \, $V$\,, \,$A_0^{}$\,, \,$T_1^{}$} ~,
\end{eqnarray}
\vspace{-4ex}
\begin{eqnarray}
\textsf{\textsl F}(s) \,\,=\,\,
\frac{r_1^{}}{1-s/m_{\rm fit}^2} \,+\, \frac{r_2^{}}{\bigl(1-s/m_{\rm fit}^2\bigr)^2}
\hspace{5ex} \mbox{for \, $A_2^{}$\,, \,$\tilde T_3^{}$} ~,
\end{eqnarray}
where \,$\tilde T_3^{}=T_2^{}+s\,T_3^{}/\bigl(m_B^2-m_V^2\bigr)$,
with \,$m_B^{}=m_{B^0}^{},m_{B_s}^{}$ and $m_V^{}=m_{K^*}^{},m_\phi^{}$.\,
The parameter values in the \,$\bar B\to\bar K^{(*)}$\, and \,$\bar B_s^{}\to\phi$\, cases
are collected in Tables~\ref{ffparameters} and~\ref{ffparameters'}, respectively.
The estimated uncertainty of each $\textsf{\textsl F}(s)$ is about
$\pm15\%$~\cite{Bobeth:2007dw,Ball:2004ye,Ball:2004rg}.

\begin{table}[ht]
\caption{Parameters for \,$B_s^{}\to\phi$\, form factors~\cite{Ball:2004rg}.
The $m_{\rm fit}^2$ and $m_R^{}$ entries are in units of $\rm GeV^2$ and GeV, respectively.}
\medskip
\begin{tabular}{|c|ccccccc|}
\hline
$\vphantom{\int_|^|}$ & $A_0^{}$ & $A_1^{}$ & $A_2^{}$ & $V$ &
$T_1^{}$ & $T_2^{}$ & $\tilde T_3^{}$ \\
\hline\hline
$~r_1^{\vphantom{|}}~$ &
   3.310    &    --   & ~~$-0.054$~~ &    1.484    &    1.303   &     --    & ~0.027~ \\
$r_{2_{\vphantom{|}}}^{\vphantom{|}}$ &
~$-2.835$~~ & ~0.308~ &   ~~0.288~   & ~$-1.049$~~ & ~$-0.954$~~ & ~~0.349~~ & ~0.321~ \\
~$m_{\rm fit_{\vphantom{|}}}^2$~ &
     31.57  &  36.54  &    ~~48.94~  &     39.52   &    46.31    &   37.21   & ~45.56~ \\
~$m_{R_{\vphantom{|}}}^{}$~ &
      5.37  &    --   &        --    &     5.41    &    5.41     &    --     &  --    \\
\hline
\end{tabular} \label{ffparameters'} \bigskip
\end{table}

\section{Squared amplitudes and decay rates\label{rates}}

The general expressions for the (double) differential decay rates of
\,$\bar B\to\bar K^{(*)}\ell^+\ell^-$\, arising from various \,$b\to s\ell^+\ell^-$\,
four-fermion operators within and beyond the SM are known in
the literature~\cite{Greub:1994pi,Deshpande:1988mg}.
Here, for completeness, we write down the specific formulas resulting from
our amplitudes of interest.

The absolute square of the \,$\bar B\to\bar K\ell^+\ell^-$\, amplitude in
Eq.~(\ref{MB2Kll}), summed over the lepton spins, is 
\begin{eqnarray} \label{M2B2Kll}
\bigl|{\cal M}\bigl(\bar B\to\bar K\ell^+\ell^-\bigr)\bigr|^2 &=&
\frac{\alpha^2 G_{\rm F}^2\,|\lambda_t^{}|^2}{\pi^2} \Bigl\{
|A|^2\, \bigl[ \bigl(t-m_\ell^2\bigr)\bigl(u-m_\ell^2\bigr)-m_B^2 m_K^2\bigr]
\nonumber \\ && \hspace*{12ex} +\;
|C|^2\, \bigl[ t\,u-\bigl(m_B^2-m_\ell^2\bigr)\bigl(m_K^2-m_\ell^2\bigr)\bigr]
\nonumber \\ && \hspace*{12ex} +\;
|D|^2\,m_\ell^2\,q^2 \,+\,
2\,{\rm Re}(C^*D)\,\bigl(m_B^2-m_K^2\bigr)m_\ell^2 \Bigr\} ~,
\end{eqnarray}
with \,$t=\bigl(p_B^{}-p_{\ell^+}^{}\bigr){}^2$\, and
\,$u=\bigl(p_B^{}-p_{\ell^-}^{}\bigr){}^2$.\,
This leads to the differential decay rate
\begin{eqnarray} \label{dGB2Kll}
\frac{d\Gamma\bigl(\bar B\to\bar K\ell^+\ell^-\bigr)}{dq^2}  &=&
\frac{\alpha^2 G_{\rm F}^2|\lambda_t^{}|^2 m_B^3 v}{2^{10}\,\pi^5}
\sqrt{\xi} \biggl\{ |A|^2 \biggl(\xi-\frac{\xi v^2}{3}\biggr)
+ |C|^2 \biggl[\xi-\frac{\xi v^2}{3}+(2+2\rho-z)\bigl(1-v^2\bigr)z \biggr]
\nonumber \\ && \hspace*{19ex} +\;
|D|^2 \bigl(1-v^2\bigr) z^2 + 2\,{\rm Re}(C^*D)\,(1-\rho)\bigl(1-v^2\bigr)z \Bigr\} ~, ~~~~~~
\end{eqnarray}
where \,$\xi=1 - 2\rho - 2z + (\rho - z)^2$,\, \,$\rho=m_K^2/m_B^2$,\,
\,$v=\sqrt{1-4m_\ell^2/q^2}$,\, and \,$z=q^2/m_B^2$.\,

The absolute square of the \,$\bar B\to\bar K^*\ell^+\ell^-$\, amplitude in
Eq.~(\ref{MB2Ksll}), summed over the $\bar K^*$ polarization and lepton spins, can be written as
\begin{eqnarray} \label{M2B2Ksll}
&& \hspace*{-3ex}
\bigl|{\cal M}\bigl(\bar B\to\bar K^*\ell^+\ell^-\bigr)\bigr|^2 \,\,=\,\,
\frac{\alpha^2G_{\rm F}^2|\lambda_t^{}|^2m_B^2}{2^4\,\pi^2\,r} \,\times
\nonumber \\ && \hspace*{8ex} \Bigl\{
\bigl(|{\cal A}|^2+|{\cal E}|^2\bigr) m_B^4 \Bigl[
\lambda\bigl(2-v^2\bigr)+\bigl(\hat t-\hat u\bigr)^2 \Bigr]r\,z
+ 4|{\cal C}|^2 r\bigl(3-v^2\bigr)z + 8|{\cal F}|^2 r v^2 z
\nonumber \\ && \hspace*{9ex} +\;
\Bigl[|{\cal C}|^2+|{\cal F}|^2 + \bigl(|{\cal D}|^2+|{\cal G}|^2\bigr)\lambda\,m_B^4
+ 2\,{\rm Re}({\cal C}^*{\cal D}+{\cal F}^*{\cal G})\,m_B^2(r+z-1)\Bigr]
\Bigl[\lambda-\bigl(\hat t-\hat u\bigr)^2\Bigr]
\nonumber \\ && \hspace*{9ex} -\;
\Bigl[ 2|{\cal E}|^2r+|{\cal G}|^2(z-2-2r) - |{\cal H}|^2 z
+ 2\,{\rm Re}({\cal G}^*{\cal H})\,(r-1)\Bigr] \lambda\,m_B^4\bigl(1-v^2\bigr)z
\nonumber \\ && \hspace*{9ex} -\;
2\,{\rm Re}({\cal F}^*{\cal G}+{\cal F}^*{\cal H})\,\lambda\,m_B^2\bigl(1-v^2\bigr)z
\,+\, 8\,{\rm Re}({\cal A}^*{\cal F}+{\cal C}^*{\cal E})\,r(t-u)z \Bigr\} ~,
\end{eqnarray}
where \,$\lambda=1 - 2r - 2z + (r - z)^2$,\, \,$r=m_{K^*}^2/m_B^2$,\,
\,$\hat t=t/m_B^2$,\, and \,$\hat u=u/m_B^2$.\,
The corresponding differential decay rate is
\begin{eqnarray} \label{dGB2Ksll}
&& \hspace*{-3ex}
\frac{d\Gamma\bigl(\bar B\to\bar K^*\ell^+\ell^-\bigr)}{dq^2} \,\,=\,\,
\frac{\alpha^2G_{\rm F}^2|\lambda_t^{}|^2}{2^{12}\,\pi^5}\,
\frac{\lambda^{3/2}\,m_B^5\,v}{3\,r}\,\times
\nonumber \\ && \hspace*{10ex} \biggl\{
\biggl[ 2|{\cal A}|^2r\,z \,+\, \frac{|{\cal C}|^2}{m_B^4}\biggl(1+\frac{12r\,z}{\lambda}\biggr) \,+\,
\lambda\,|{\cal D}|^2 \biggr] \bigl(3-v^2\bigr)
\,+\, 4\,|{\cal E}|^2 r\, v^2\, z \,+\, 3\,|{\cal H}|^2 \bigl(1-v^2\bigr) z^2
\nonumber \\ && \hspace*{11ex} +\;
\frac{|{\cal F}|^2}{m_B^4} \biggl(3-v^2+\frac{24r\,v^2\,z}{\lambda}\biggr)
\,+\, |{\cal G}|^2 \Bigl[ (1-r)^2 \bigl(3-v^2\bigr)+2(z-2-2r)v^2z\Bigr]
\nonumber \\ && \hspace*{11ex} +\;
\frac{2\,{\rm Re}({\cal C}^*{\cal D})}{m_B^2} (r+z-1) \bigl(3-v^2\bigr)
\,+\, \frac{2\,{\rm Re}({\cal F}^*{\cal G})}{m_B^2} \Bigl[(1-r)\bigl(v^2-3\bigr)+2v^2\,z\Bigr]
\nonumber \\ && \hspace*{11ex} +\;
6\biggl[ \frac{{\rm Re}({\cal F}^*{\cal H})}{m_B^2}
+ {\rm Re}({\cal G}^*{\cal H})\,(r-1) \biggr] \bigl(v^2-1\bigr)z \biggr\} ~.
\end{eqnarray}
From Eqs.~(\ref{d2GB2Ksll}) and~(\ref{M2B2Ksll}), we can derive the $\bar K^*$
longitudinal-polarization fraction
\begin{eqnarray}
F_L^{}  &=&
\frac{\alpha^2G_{\rm F}^2\,|\lambda_t^{}|^2\,\lambda^{3/2}\,m_B^5\,v}
{2^{11}\,3\pi^5\,r\,\Gamma'\bigl(\bar B\to\bar K^*\ell^+\ell^-\bigr)} \Biggl\{
\Biggl[ |{\cal A}|^2 r+ \frac{|{\cal H}|^2 z}{2}
- \frac{{\rm Re}({\cal F}^*{\cal H})}{m_B^2} + {\rm Re}({\cal G}^*{\cal H})\, (1-r) \Biggr]
\bigl(1-v^2\bigr)z
\nonumber \\ && \hspace*{28ex} +\;
\frac{|{\cal C}|^2}{m_B^4} \Biggl[ 2r\bigl(3-v^2\bigr)\frac{z}{\lambda} + \frac{1+v^2}{2} \Biggr]
\nonumber \\ && \hspace*{28ex} +\;
\Biggl[ \frac{|{\cal D}|^2 \lambda}{2} + \frac{{\rm Re}({\cal C}^*{\cal D})}{m_B^2}(r+z-1)
\Biggr] \bigl(1+v^2\bigr)
\nonumber \\ && \hspace*{28ex} +\;
\frac{|{\cal F}|^2}{m_B^4} \biggl( \frac{4 r v^2 z}{\lambda} + \frac{1+v^2}{2} \biggr) +
|{\cal G}|^2 \Biggl[ \lambda v^2 + (1-r)^2\frac{1-v^2}{2} \Biggr]
\nonumber \\ && \hspace*{28ex} +\;
\frac{{\rm Re}({\cal F}^*{\cal G})}{m_B^2}\left[r - 1 + (r+2z-1)v^2\right] \Biggr\}
\end{eqnarray}
and lepton forward-backward asymmetry
\begin{eqnarray}
A_{\rm FB}^{} &=&
\frac{1}{\Gamma'\bigl(\bar B\to\bar K^*\ell^+\ell^-\bigr)}
\int_{-1}^1d c_\theta^{}\;
\frac{d^2\Gamma\bigl(\bar B\to\bar K^*\ell^+\ell^-\bigr)}{dq^2\,dc_\theta^{}}\;{\rm sgn}\,c_\theta^{}
\nonumber \\ &=&
\frac{-\alpha^2G_{\rm F}^2\,|\lambda_t^{}|^2\,\lambda\,m_B^3\,v^2\,z}
{2^{10}\,\pi^5\,\Gamma'\bigl(\bar B\to\bar K^*\ell^+\ell^-\bigr)}\,
{\rm Re}({\cal A}^*{\cal F}+{\cal C}^*{\cal E}) ~.
\end{eqnarray}
where
\,$\Gamma'\bigl(\bar B\to\bar K^*\ell^+\ell^-\bigr)\equiv
d\Gamma\bigl(\bar B\to\bar K^*\ell^+\ell^-\bigr)/dq^2$\,
and \,$c_\theta^{}\equiv\cos\theta$.\,

We remark that the \,$t-u$\, term in the last line of Eq.~(\ref{M2B2Ksll}) contains $A_{\rm FB}^{}$
for  \,$\bar B\to\bar K^*\ell^+\ell^-$.\,
Since such a term is absent in Eq.~(\ref{M2B2Kll}), there is no $A_{\rm FB}^{}$ for
\,$\bar B\to\bar K\ell^+\ell^-$\, from the SM or from the $X$ contributions under consideration.


\begin{thebibliography}{0}

\bibitem{Abazov:2010hv}
  V.M.~Abazov {\it et al.}  [D0 Collaboration],
  arXiv:1005.2757 [hep-ex];
  arXiv:1007.0395 [hep-ex].

\bibitem{Oh:2010vc}
  S.~Oh and J.~Tandean,
  Phys.\ Lett.\  B {\bf 697}, 41 (2011)
  [arXiv:1008.2153 [hep-ph]].

\bibitem{Gninenko:2001hx}
  S.N.~Gninenko and N.V.~Krasnikov,
  Phys.\ Lett.\  B {\bf 513}, 119 (2001)
  [arXiv:hep-ph/0102222];
  C.~Boehm,
  Phys.\ Rev.\  D {\bf 70}, 055007 (2004)
  [arXiv:hep-ph/0405240].

\bibitem{Hooper:2007jr}
  D.~Hooper,
  Phys.\ Rev.\  D {\bf 75}, 123001 (2007)
  [arXiv:hep-ph/0701194];
  P.~Fayet,
  Phys.\ Rev.\  D {\bf 75}, 115017 (2007)
  [arXiv:hep-ph/0702176].

\bibitem{Foot:1994vd}
  R.~Foot, X.G.~He, H.~Lew, and R.R.~Volkas,
  Phys.\ Rev.\  D {\bf 50}, 4571 (1994)
  [arXiv:hep-ph/9401250];
  P.f.~Yin, J.~Liu, and S.h.~Zhu,
  Phys.\ Lett.\  B {\bf 679}, 362 (2009)
  [arXiv:0904.4644 [hep-ph]].

\bibitem{He:2005we}
  X.G.~He, J.~Tandean, and G.~Valencia,
  Phys.\ Lett.\ B {\bf 631}, 100 (2005)
  [arXiv:hep-ph/0509041];

\bibitem{Chen:2007uv}
  C.H.~Chen, C.Q.~Geng, and C.W.~Kao,
  Phys.\ Lett.\  B {\bf 663}, 400 (2008)
  [arXiv:0708.0937 [hep-ph]];

\bibitem{Oh:2009fm}
 S.~Oh and J.~Tandean,
  JHEP {\bf 1001}, 022 (2010)
  [arXiv:0910.2969 [hep-ph]].

\bibitem{Pospelov:2008zw}
  M.~Pospelov,
  Phys.\ Rev.\  D {\bf 80}, 095002 (2009)
  [arXiv:0811.1030 [hep-ph]].
  M.~Reece and L.T.~Wang,
  JHEP {\bf 0907}, 051 (2009)
  [arXiv:0904.1743 [hep-ph]];

\bibitem{Aubert:2004it}
  B.~Aubert {\it et al.}  [BABAR Collaboration],
  Phys.\ Rev.\ Lett.\  {\bf 93}, 081802 (2004)
  [arXiv:hep-ex/0404006].

\bibitem{Iwasaki:2005sy}
  M.~Iwasaki {\it et al.}  [Belle Collaboration],
  Phys.\ Rev.\  D {\bf 72}, 092005 (2005)
  [arXiv:hep-ex/0503044].

\bibitem{Aubert:2008ju}
  B.~Aubert {\it et al.}  [BABAR Collaboration],
  Phys.\ Rev.\  D {\bf 79}, 031102 (2009)
  [arXiv:0804.4412 [hep-ex]];
  Phys.\ Rev.\ Lett.\  {\bf 102}, 091803 (2009)
  [arXiv:0807.4119 [hep-ex]].

\bibitem{Wei:2009zv}
  J.T.~Wei {\it et al.}  [BELLE Collaboration],
  Phys.\ Rev.\ Lett.\  {\bf 103}, 171801 (2009)
  [arXiv:0904.0770 [hep-ex]].

\bibitem{Aaltonen:2011cn}
  T.~Aaltonen {\it et al.}  [CDF Collaboration],
  arXiv:1101.1028 [hep-ex].

\bibitem{Greub:1994pi}
  C.~Greub, A.~Ioannisian, and D.~Wyler,
  Phys.\ Lett.\  B {\bf 346}, 149 (1995)
  [arXiv:hep-ph/9408382];
  S.~Fukae, C.S.~Kim, T.~Morozumi, and T.~Yoshikawa,
  Phys.\ Rev.\  D {\bf 59}, 074013 (1999)
  [arXiv:hep-ph/9807254];
  T.M.~Aliev, C.S.~Kim, and Y.G.~Kim,
  Phys.\ Rev.\  D {\bf 62}, 014026 (2000)
  [arXiv:hep-ph/9910501];
  C.W.~Chiang, R.H.~Li, and C.D.~Lu,
  arXiv:0911.2399 [hep-ph];
  Q.~Chang, X.Q.~Li, and Y.D.~Yang,
  JHEP {\bf 1004}, 052 (2010)
  [arXiv:1002.2758 [hep-ph]];
  A.K.~Alok, A.~Datta, A.~Dighe, M.~Duraisamy, D.~Ghosh, D.~London, and S.U.~Sankar,
  arXiv:1008.2367 [hep-ph].

\bibitem{Altmannshofer:2008dz}
  W.~Altmannshofer, P.~Ball, A.~Bharucha, A.J.~Buras, D.M.~Straub, and M.~Wick,
  JHEP {\bf 0901}, 019 (2009)
  [arXiv:0811.1214 [hep-ph]].

\bibitem{Deshpande:1988mg}
  N.G.~Deshpande and J.~Trampetic,
  Phys.\ Rev.\ Lett.\  {\bf 60}, 2583 (1988).
  B.~Grinstein, M.J.~Savage, and M.B.~Wise,
  Nucl.\ Phys.\  B {\bf 319}, 271 (1989);
  W.~Jaus and D.~Wyler,
  Phys.\ Rev.\  D {\bf 41}, 3405 (1990);
  A.~Ali, T.~Mannel, and T.~Morozumi,
  Phys.\ Lett.\  B {\bf 273}, 505 (1991).

\bibitem{Burdman:1995ks}
  G.~Burdman,
  Phys.\ Rev.\  D {\bf 52}, 6400 (1995)
  [arXiv:hep-ph/9505352];
  A.~Ali, P.~Ball, L.T.~Handoko, and G.~Hiller,
  Phys.\ Rev.\  D {\bf 61}, 074024 (2000)
  [arXiv:hep-ph/9910221];
  M.~Beneke, T.~Feldmann, and D.~Seidel,
  Nucl.\ Phys.\  B {\bf 612}, 25 (2001)
  [arXiv:hep-ph/0106067].

\bibitem{Kruger:2005ep}
  F.~Kruger and J.~Matias,
  Phys.\ Rev.\  D {\bf 71}, 094009 (2005)
  [arXiv:hep-ph/0502060].

\bibitem{Leveille:1977rc}
  J.P.~Leveille,
  Nucl.\ Phys.\  B {\bf 137}, 63 (1978).

\bibitem{Huber:2005ig}
  T.~Huber, E.~Lunghi, M.~Misiak, and D.~Wyler,
  Nucl.\ Phys.\  B {\bf 740}, 105 (2006)
  [arXiv:hep-ph/0512066];
  T.~Huber, T.~Hurth, and E.~Lunghi,
  Nucl.\ Phys.\  B {\bf 802}, 40 (2008)
  [arXiv:0712.3009 [hep-ph]].

\bibitem{pdg}
K. Nakamura {\it et al.}  [Particle Data Group],
  J.~Phys. G {\bf 37}, 075021 (2010).

\bibitem{ckmfit}
CKMfitter, http://ckmfitter.in2p3.fr.

\bibitem{Chen:2005ht}
  C.H.~Chen and H.N.~Li,
  Phys.\ Rev.\  D {\bf 71}, 114008 (2005)
  [arXiv:hep-ph/0504020].

\bibitem{Gao:2006yu}
  Y.J.~Gao, C.~Meng, and K.T.~Chao,
  Eur.\ Phys.\ J.\  A {\bf 28}, 361 (2006)
  [arXiv:hep-ph/0606044].

\bibitem{Ball:2004ye}
  P.~Ball and R.~Zwicky,
  Phys.\ Rev.\  D {\bf 71}, 014015 (2005)
  [arXiv:hep-ph/0406232].

\bibitem{Ball:2004rg}
  P.~Ball and R.~Zwicky,
  Phys.\ Rev.\  D {\bf 71}, 014029 (2005)
  [arXiv:hep-ph/0412079].

\bibitem{Bobeth:2007dw}
  C.~Bobeth, G.~Hiller and G.~Piranishvili,
  JHEP {\bf 0712}, 040 (2007)
  [arXiv:0709.4174 [hep-ph]];
  C.~Bobeth, G.~Hiller and D.~van Dyk,
  JHEP {\bf 1007}, 098 (2010)
  [arXiv:1006.5013 [hep-ph]].

\bibitem{Jegerlehner:2009ry}
  F.~Jegerlehner and A.~Nyffeler,
  Phys.\ Rept.\  {\bf 477}, 1 (2009)
  [arXiv:0902.3360 [hep-ph]].

\bibitem{Onogi:2006km}
  T.~Onogi,
  PoS {\bf LAT2006}, 017 (2006)
  [arXiv:hep-lat/0610115].

\bibitem{Geng:2003su}
  C.Q.~Geng and C.C.~Liu,
  J.\ Phys.\ G {\bf 29}, 1103 (2003)
  [arXiv:hep-ph/0303246].

\end{thebibliography}
\end{document}